\documentclass[journal=jctcce,manuscript=article]{achemso}

\usepackage[version=3]{mhchem} %
\usepackage[dvipsnames]{xcolor}
\usepackage{amsfonts} 
\usepackage{caption}
\usepackage{subcaption}
\usepackage{hyperref}
\usepackage{cleveref}
\usepackage{tikz}
\usepackage{footmisc}
\usepackage{pdfpages}
\usetikzlibrary{shapes.geometric, arrows, positioning}

\mciteErrorOnUnknownfalse %

\DeclareMathOperator*{\argmin}{arg\,min}

\newcommand{\vk}{\mathbf{k}}
\newcommand{\vq}{\mathbf{q}}
\newcommand{\vG}{\mathbf{G}}
\newcommand{\vx}{\mathbf{x}}

\newcommand{\vzero}{\mathbf{0}}
\newcommand{\primedsum}{\sideset{}{'}\sum}
\newcommand{\sigmafit}{\sigma_{\text{fit}}}

\newcommand{\Nocc}{N_{\text{occ}}}

\newcommand{\Nk}{N_{\vk}}

\setkeys{acs}{maxauthors = 10}

\newcommand{\firstpassrev}[1]{{#1}}
\newcommand{\REV}[1]{#1}

\tikzstyle{process} = [rectangle, draw, text centered, minimum width=8cm, minimum height=1.2cm]
\tikzstyle{highlight} = [rectangle, draw=red, line width=2.5pt, text centered, minimum width=8cm, minimum height=1.2cm]
\tikzstyle{arrow} = [thick, ->, >=stealth]

\author{Stephen Jon Quiton}
\author{Juan D. F. Pottecher}

\affiliation{Department of Chemistry, University of California, Berkeley, California 94720, United States}

\author{Xin Xing}
\affiliation{Department of Mathematics, University of California, Berkeley, California 94720, United States}

\author{Martin Head-Gordon}
\email{mhg@cchem.berkeley.edu}
\affiliation{Department of Chemistry, University of California, Berkeley, California 94720, United States}

\author{Lin Lin}
\email{linlin@math.berkeley.edu}
\affiliation{Department of Mathematics, University of California, Berkeley, California 94720, United States}
\alsoaffiliation{Applied Mathematics and Computational Research Division, Lawrence Berkeley National Laboratory, Berkeley, California 94720, United States}

\title[]{Optimized auxiliary functions for robust mitigation of finite-size errors in periodic hybrid density functional theory}

\keywords{American Chemical Society, \LaTeX}

\begin{document}

\begin{tocentry}

\includegraphics[width=\textwidth]{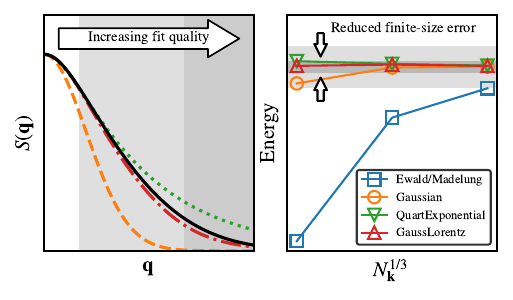}

\end{tocentry}

\begin{abstract}

When calculating properties of periodic systems at the thermodynamic limit (TDL), the dominant source of finite size error (FSE) arises from the long-range Coulomb interaction, and can manifest as a slowly converging quadrature error when approximating an integral in the reciprocal space by a finite sum. The singularity subtraction (SS) method offers a systematic approach for reducing this quadrature error and thus the FSE. In this work, we first investigate the performance of the SS method in the simplest setting,  aiming at reducing the FSE in exact exchange calculations by subtracting the Coulomb contribution with a single, adjustable Gaussian auxiliary function. We demonstrate that a simple fitting method can robustly estimate the optimal Gaussian width and leads to rapid convergence toward the TDL. Furthermore, we suggest new forms of the auxiliary function, whose optimal parameters could also be determined through least-squares fitting. For a range of semiconductors and insulators, the proposed auxiliary functions achieve robust, millihartree-level accuracy in hybrid density functional theory calculations, including cases with sparse $\vk$-meshes and large basis sets.%

\end{abstract}

\section{Introduction}

Hybrid Density Functional Theory (DFT) calculations are becoming increasingly popular for computing the properties of molecules and solids~\cite{stadeleExactExchangeKohnSham1999,coraPerformanceHybridDensity2004,vandewalleFirstprinciplesCalculationsDefects2004,dovesiPeriodicHartreeFockMethod2000,dovesiInitioQuantumSimulation2005,hafnerAbinitioSimulationsMaterials2008,freysoldtElectrostaticInteractionsCharged2011,mattioliPhotocatalyticPhotovoltaicProperties2014,refaely-abramsonSolidstateOpticalAbsorption2015,chenFirstprinciplesDeterminationDefect2015,paierHybridDensityFunctionals2016,liaoDensityFunctionalTheory2022}. While pure DFT functionals---including the local density approximation (LDA) and the generalized gradient approximation (GGA)---have the advantage in terms of computational cost scaling, they often produce critical physical inaccuracies, including underestimation of band-gaps  and over-delocalized densities, which result primarily from  self-interaction errors\cite{stadeleExactExchangeKohnSham1999,dovesiPeriodicHartreeFockMethod2000,heydEnergyBandGaps2005,schimkaImprovedHybridFunctional2011}. Including a portion of Fock exchange to cancel out these self-interaction errors drastically improves the accuracy while scaling only about a power higher in computational cost as a function of the number of atomic orbitals $N_{\text{AO}}$ in the system, making hybrid DFT functionals still scale more favorably than wavefunction-based correlated methods like Coupled Cluster or M{\o}ller-Plesset perturbation theory.

Finite-size errors (FSEs), which arise from the $\vk$-point discretization of integrals over reciprocal space, are fundamental to periodic quantum chemistry calculations and present a third and orthogonal source of error to basis set incompleteness (BSIE) and correlation errors encountered in molecular calculations. Equivalently in the real-space representation, FSEs correspond to the deviation in calculations using finite computational unit cells treated with periodic boundary conditions versus results at the thermodynamic limit (TDL). The FSE for the Hartree-Fock (HF) energy scales as $\mathcal{O}(N_{\vk}^{-1/3})$ in the absence of correction schemes, where  $\Nk$ is the number of $\vk$-points sampled in the FBZ. This error originates primarily from excluded divergent term in the exchange operator and can be reduced to $\mathcal{O}(N_{\vk}^{-1})$ with the widely-used Madelung constant correction\cite{xingUnifiedAnalysisFinitesize2024}. 
There has been extensive work in reducing the FSE in the exchange operator, such as the truncated Coulomb potential~\cite{spencerEfficientCalculationExact2008,sundararamanRegularizationCoulombSingularity2013}, mean Coulomb potential~\cite{schaferSamplingReciprocalCoulomb2024}, and the staggered mesh method~\cite{quitonStaggeredMeshMethod2024}, to name a few. These methods are also applicable to hybrid DFT with long-range Coulomb interactions in the exchange term, such as the PBE0 functional~\cite{perdewGeneralizedGradientApproximation1996,ernzerhofAssessmentPerdewBurke1999,adamoReliableDensityFunctional1999}.

This work focuses on a class of methods for reducing the finite-size error (FSE) in exact exchange energy calculations based on singularity subtraction (also called the auxiliary function method), which we refer to as ExxSS. The singularity subtraction method was first pioneered by Gygi and Baldereschi~\cite{gygiSelfconsistentHartreeFockScreenedexchange1986} to mitigate FSEs in face-centered cubic (fcc) structures, and was later generalized to other crystal symmetries by Wenzien and collaborators~\cite{wenzienEfficientQuasiparticleBandstructure1995}. Notably, the Madelung constant correction can be interpreted as a leading-order special case of singularity subtraction~\cite{xingUnifiedAnalysisFinitesize2024}.
A widely adopted form of ExxSS employs a Gaussian auxiliary function, first introduced by Massidda et al.~\cite{massiddaHartreeFockLAPWApproach1993}, where the Gaussian width is chosen to approximate the diameter of the first Brillouin zone (FBZ). While the Madelung correction can be interpreted as a constant approximation to the structure factor\footnote{Not to be confused with the structure factor in the scattering context.} $S(\mathbf{q})$, the Gaussian auxiliary function provides a more accurate approximation to its shape, and has been used in many studies~\cite{sorouriAccurateEfficientMethod2006,ducheminScalableAccurateAlgorithm2010,kawashimaSingularityCorrectionLongRangeCorrected2017,zavickisAdaptivelyCompressedExchange2022,broqvistHybridfunctionalCalculationsPlanewave2009,nguyenEfficientCalculationExact2009,holzwarthAnalysisNumericalMethods2011,liElectronicLevelsElectrical2011,chenFirstprinciplesDeterminationDefect2015,bylaskaCorrespondingOrbitalsDerived2018}. However, the parameters of the auxiliary function are typically chosen based on simple system characteristics, such as the simulation cell geometry or kinetic energy cutoff, without leveraging finer-grained information from the structure factor.

In this work, we systematically investigate the selection of auxiliary functions within the ExxSS framework. We begin by examining the Gaussian auxiliary function and demonstrate that its optimal width parameter, $\sigma$, can be obtained via a least-squares fit to the structure factor in reciprocal space. We denote this optimal value by $\sigma_{\mathrm{fit}}$, and find that it strongly depends on the maximum distance from the $\Gamma$-point (i.e., $\mathbf{q}=0$) among the quadrature nodes used in the fitting. Numerical experiments show that fitting to the $\mathbf{q}$-points nearest the $\Gamma$-point yields the most effective $\sigma_{\mathrm{fit}}$, suggesting that accurately capturing the curvature of the structure factor near the origin is key to reducing FSEs. %

Building on the success of the Gaussian-based ExxSS with optimized parameters, we may ask: is Gaussian the optimal choice of auxiliary function? One observation is that when approaching the complete basis set (CBS) limit, the decay rate of the structure factor $S(\mathbf{q})$ is generally much slower than that of  Gaussian functions. This motivates the exploration of alternative auxiliary functions that better capture both the near-origin behavior and the mid-to-long range behavior of $S(\mathbf{q})$, thereby addressing both the asymptotic behavior and the prefactor of the FSE.

We propose new classes of auxiliary functions, including Quartic Exponential and Gaussian-Lorentzian hybrids, which combine accurate curvature matching near the $\Gamma$-point with slower-decaying tails. We find that these new functions not only significantly outperform both the Madelung correction and ExxSS with the optimized Gaussian in reducing FSEs, but are also more robust with respect to fitting hyper-parameters. Furthermore, the additional cost is small compared to the cost of the self-consistent field (SCF) procedure and the construction of Cholesky vectors for Gaussian Density Fitting (GDF)~\cite{yeFastPeriodicGaussian2021}.
Based on our numerical results, we recommend the Quartic Exponential auxiliary function, which involves three fitting parameters, as a robust and accurate alternative within the ExxSS framework.

The rest of the paper is as follows. We first outline the theory behind the FSE scaling for the Fock exchange energy and give an overview of how singularity subtraction can be used to reduce it. Then, we investigate the use of various auxiliary functions, first showing how to obtain the optimal parameters for the popular Gaussian auxiliary function and then demonstrating that more robust auxiliary functions can still further reduce the FSE. Finally, we test a select few auxiliary functions for their ability to recover the TDL for energies and bulk properties.

\section{Theory}
We use the Bloch orbital framework, denoting atomic orbitals (AOs) by $\mu,\nu,\lambda,\sigma$, occupied crystalline/molecular orbitals (MOs) by $i,j,k,l$, and virtual MOs by $a,b,c,d$. Then the Bloch orbital with band index $\mu$ and crystal momentum $\vk$ is defined as  
\begin{equation}
    \psi_{\mu\mathbf{k}}(\mathbf{r})=\frac{1}{\sqrt{\Nk}} \sum_{\mathbf{R}\in\mathbb{L}} \mathrm{e}^{\mathrm{i} \mathbf{k} \cdot \mathbf{R}} \phi_{\mu}(\mathbf{r}-\mathbf{R}),
\end{equation}
where $\mathbf{R}$ is a vector in the real space Bravais-lattice $\mathbb{L}$,  and $\phi_{\mu}$ is an atomic orbital.
The standard approach in sampling the reciprocal space is to use a uniform grid called the  Monkhorst-Pack (MP) mesh \cite{monkhorstSpecialPointsBrillouinzone1976}. We denote by $N_\mathrm{k}$ the total number of $\vk$ points in the mesh, and denote the unit cell by $\Omega$ with volume $\left|\Omega\right|$. We also let $|\Omega^s|:=N_{k} |\Omega|$. Next, define the $\vk$-dependent AO basis function $\phi_{\mu\mathbf{k}}(\mathbf{r})=e^{-\mathrm{i}\mathbf{k}\cdot\mathbf{r}}\psi_{\mu\mathbf{k}}(\mathbf{r})$. Then, the product of two such basis functions is defined as  

\begin{equation}
\rho_{\REV{\mu\mathbf{k}_{\mu}\nu\mathbf{k}_{\nu}}}(\mathbf{r})=\phi_{\mu\mathbf{k}_{\mu}}^{*}\left(\mathbf{r}\right)\phi_{\nu\mathbf{k}_{\nu}}\left(\mathbf{r}\right):=\frac{1}{|\Omega|}\sum_{\mathbf{G}\in\mathbb{L}^{*}}\hat{\rho}_{\REV{\mu\mathbf{k}_{\mu}\nu\mathbf{k}_{\nu}}}(\mathbf{G})e^{\mathrm{i}\mathbf{G}\cdot\mathbf{r}},\end{equation}
where \REV{$\hat{\rho}_{\mu{\mathbf{k}_{\mu}}\nu{\mathbf{k}_{\nu}}}(\mathbf{G})$} is the Fourier transform of the AO product and \REV{$\mathbb{L}^*$ is the reciprocal space lattice associated with $\mathbb{L}$}. With this definition, two-electron integrals in the Bloch orbital basis \REV{(using AOs for the periodic component)} can be expressed as  
\begin{align}
\left\langle\mu\mathbf{k}_{\mu},\nu\mathbf{k}_{\nu}\mid\lambda\mathbf{k}_{\lambda},\sigma\mathbf{k}_{\sigma}\right\rangle &=\frac{1}{\left|\Omega^{s}\right|}\sideset{}{'}\sum_{\mathbf{G}\in\mathbb{L}^{*}}\hat{\rho}_{\REV{\mu\mathbf{k}_{\mu}\lambda\mathbf{k}_{\lambda}}}(\mathbf{G})\hat{V}_{\mathrm{coul}}\left(\mathbf{q}+\mathbf{G}\right)\hat{\rho}_{\nu\mathbf{k}_{\nu}\sigma\mathbf{k}_{\sigma}}\left(\mathbf{G}_{\mathbf{k}_{\mu},\mathbf{k}_{\nu}}^{\mathbf{k_{\lambda}},\mathbf{k}_{\sigma}}-\mathbf{G}\right).\label{eq:eri-reciprocal}
\end{align}
Here, $\mathbf{G}_{\mathbf{k}_{\mu},\mathbf{k}_{\nu}}^{\mathbf{k}_{\lambda},\mathbf{k}_{\sigma}}:=\mathbf{k}_{\mu}+\mathbf{k}_{\nu}-\mathbf{k}_{\lambda}-\mathbf{k}_{\sigma}\in \mathbb{L}^*$ by crystal momentum conservation,  $\mathbf{q}=\mathbf{k}_\lambda-\mathbf{k}_\nu$ is the momentum transfer vector, $\sum'$ indicates that the (singular) term where 
$\mathbf{q}+\mathbf{G}=\mathbf{0}$ is excluded, and \firstpassrev{$\hat{V}_{\text{coul}}(\vq+\vG)=4\pi/|\vq+\vG|^2$ is the Coulomb kernel in reciprocal space}.

If we define $C_{\mu i}^\mathbf{k}$ to be the coefficient matrix for the crystalline orbitals (using MOs for the periodic component),
\begin{align*}
\REV{\psi_{i\mathbf{k}}(\mathbf{r})}=\sum_{\mu\in \mathrm{AO}}C_{\mu i}^{\mathbf{k}}\REV{\phi_{\mu\mathbf{k}}(\mathbf{r})},
\end{align*}
then the density matrix at that same \textbf{k}, $P^\mathbf{k}$, is defined as
\begin{equation}
    P_{\mu \nu}^{\mathbf{k}}=\sum_{i \in \mathrm{occ}} C_{\mu i}^{\mathbf{k}}\left(C_{\nu i}^{\mathbf{k}}\right)^*.
\end{equation}

\subsection{Finite-size Error Scaling of Fock Exchange Energy}
\firstpassrev{With a uniform reciprocal-space mesh $\mathcal{K}$ (which converges to $\Omega$ as $\Nk$ goes to infinity)}, the Fock exchange energy per unit cell is computed as 
\begin{equation}\label{eqn:exchange_nk}
E_{\mathrm{X}}^{N_{\vk}}=\dfrac{1}{N_{\vk}}\left(-\frac{1}{2}\sum_{i,j}\sum_{\vk_{i},\vk_{j}\in\mathcal{K}}\REV{\left\langle i\vk_{i},j\vk_{j}|j\vk_{j},i\vk_{i}\right\rangle}\right).
\end{equation}
Expressed in terms of AO products and then in reciprocal space yields 
\begin{align}
    E_{\mathrm{X}}^{N_{\mathbf{k}}} & =-\frac{1}{2N_{\mathbf{k}}} \sum_{\mathbf{k}_{i} \in \mathcal{K}} \sum_{\mathbf{q} \in \mathcal{K}_{\mathbf{q}}} \sum_{i j}\left\langle i \mathbf{k}_{i}, j\left(\mathbf{k}_{i}+\mathbf{q}\right) \mid j\left(\mathbf{k}_{i}+\mathbf{q}\right), i \mathbf{k}_{i}\right\rangle\nonumber  \\
    & =-\frac{1}{2N_{\mathbf{k}}} \sum_{\mathbf{k}_{i} \in \mathcal{K}} \sum_{\mathbf{q} \in \mathcal{K}_{\mathbf{q}}} \sum_{i j} \frac{1}{|\Omega| N_{\mathbf{k}}} \primedsum_{\mathbf{G} \in \mathbb{L}^{*}}\hat{V}_{\mathrm{coul}}\left(\mathbf{q}+\mathbf{G}\right)\left|\hat{\varrho}_{i \mathbf{k}_{i}, j\left(\mathbf{k}_{i}+\mathbf{q}\right)}(\mathbf{G})\right|^{2}\nonumber  \\
    & = -\frac{\left|\Omega^{*}\right|}{16\pi^3N_{\mathbf{k}}}\sum_{\mathbf{q} \in \mathcal{K}_{\mathbf{q}}}\primedsum_{\mathbf{G} \in \mathbb{L}^{*}} \hat{V}_\mathrm{coul}(\mathbf{q}+\mathbf{G}) S\left(\mathbf{q}+\mathbf{G}\right),\label{eq:exchange_nk}
\end{align}
where $\mathcal{K}_\mathbf{q}$ is a uniform mesh in the first Brillouin zone , $\Omega^*$ (which has volume $\left|\Omega^{*}\right|$) that contains the $\Gamma$-point and has the same size as $\mathcal{K}$. 
The quantity contracted with the Coulomb kernel in reciprocal space in eq \eqref{eq:exchange_nk} is the structure factor, defined as:
\begin{equation}
S(\vq+\vG)=\frac{1}{N_{\mathbf{k}}} \sum_{\mathbf{k}_{i} \in \mathcal{K}} \sum_{i j}\left|\hat{\varrho}_{i \mathbf{k}_{i}, j\left(\mathbf{k}_{i}+\mathbf{q}\right)}(\mathbf{G})\right|^{2}.\label{eq:structure_factor}
\end{equation}
$S(\mathbf{q}+\mathbf{G})$ plays an important role in the following discussion. The prefactor in eq \eqref{eq:exchange_nk} is chosen so that $S(\mathbf{0})=\Nocc$, where $\Nocc$ is the number of electrons in the unit cell.

To approach the TDL, $\mathcal{K}$ must converge to $\Omega^*$ and $N_\vk \rightarrow\infty$. The energy computed in eq \eqref{eq:exchange_nk} thus converges to %
\begin{equation}
    E_{X}^{\text{TDL}}
    =
    \lim_{N_\vk\rightarrow\infty} E_X^{N_\vk}
    =-\frac{1}{16\pi^3}\int_{\Omega^{*}} \mathrm{~d} \mathbf{q} \sum_{\mathbf{G} \in \mathbb{L}^{*}} \hat{V}_\mathrm{coul}(\mathbf{q}+\mathbf{G}) S^\mathrm{TDL}(\mathbf{q}+\mathbf{G}),\label{eq:exchange_tdl}
  \end{equation}
where the structure factor at the TDL is defined as 
\begin{equation}
S^\mathrm{TDL}(\vq+\vG)=\frac{1}{|\Omega^{*}|} \int_{\Omega^{*}} \mathrm{d}\vk_i \sum_{i j}\left|\hat{\varrho}_{i \mathbf{k}_{i}, j\left(\mathbf{k}_{i}+\mathbf{q}\right)}(\mathbf{G})\right|^{2}.\label{eq:sqg_tdl}
\end{equation}
The calculation expressed in eq \eqref{eq:exchange_nk} is essentially a discretization of the value at the TDL expressed in eq \eqref{eq:exchange_tdl}; this discretization results in the finite-size error, which scales, according to ref ~\citenum{xingUnifiedAnalysisFinitesize2024}, as
\[
    \left| E_X^\text{TDL} - E_X^{N_\vk}\right| = \mathcal{O}\left(N_\vk^{-\frac13}\right).
\]

\subsection{Overview of Singularity Subtraction (SS)}

One way to reduce the FSE in exact exchange is through the singularity subtraction (SS) technique\cite{aliabadiTaylorExpansionsSingular1985,davisMethodsNumericalIntegration1975}, which involves adding and subtracting a function that has the same singularity shape as $\hat{V}_{\mathrm{coul}}(\mathbf{G})$ to accelerate the convergence of the quadrature.\cite{gygiSelfconsistentHartreeFockScreenedexchange1986,massiddaHartreeFockLAPWApproach1993,broqvistHybridfunctionalCalculationsPlanewave2009,xingUnifiedAnalysisFinitesize2024}. Specifically, consider a scalar-valued function $f(\vx)$ which is singular at the origin $\vx=\vzero$. One method to compute its integral with a straightforward quadrature scheme is to simply exclude the singular point; that is:
\begin{align}
    \int_{\mathbb{R}^d}f(\vx)\mathrm{d}\vx\approx \primedsum_{i=1}^M  w(\vx_i) f(\vx_i),\label{eq:regular_quadrature}
\end{align}
where again the primed sum means the singular point is excluded, and $w(\vx_i)$ is the quadrature weight at the node $i$. This approach, however, typically exhibits slow convergence towards the true value of the integral as the number of quadrature points increases, i.e., 
\begin{align}
    \left|\int_{\mathbb{R}^d}f(\vx)\mathrm{d}\vx-\primedsum_{i=1}^M  w(\vx_i) f(\vx_i)\right|=\mathcal{O}(M^{-\alpha}), \label{eq:fse_decomposition}
\end{align}
for some small exponent $\alpha$.

The idea of the SS method is to add and subtract an \firstpassrev{auxiliary function} $g(\vx)$, resulting in the following splitting of the integrand %
\begin{align}\label{eq:singularity-subtraction-general}
\int_{\mathbb{R}^d}f(\vx)\mathrm{d}\vx=\int_{\mathbb{R}^d}(f(\vx)-g(\vx))\mathrm{d}\vx + \int_{\mathbb{R}^d}g(\vx)\mathrm{d}\vx \approx \primedsum_i  w(\vx_i)(f(\vx_i)-g(\vx_i))+\int_{\mathbb{R}^d}g(\vx)\mathrm{d}\vx,
\end{align}
Here, $g(\vx)$ should be chosen so that the second term (the integral involving $g$) has a closed-form expression or is easily computable, and $g(\vx)$ can match certain low orders of the Taylor expansion of $f(\vx)$ at $\vx = 0$. Then the primed sum is identical to a regular sum in the quadrature term $\primedsum_i w(\vx_i)(f(\vx_i) - g(\vx_i))=\sum_i w(\vx_i)(f(\vx_i) - g(\vx_i))$. This also improves the convergence rate of the quadrature, yielding a larger exponent $\alpha$ in eq \eqref{eq:fse_decomposition}. If additionally, $g(\vx)$ also closely matches $f(\vx)$ across the whole domain (not just at the singularity), then $\sum_i  w(\vx_i)(f(\vx_i)-g(\vx_i))$ will be small. This can lead to a reduced error prefactor in  eq \eqref{eq:fse_decomposition}, which can be particularly effective when a small number of quadrature points are used.   In order to obtain practical SS procedures in the context of exchange energy calculations, where oftentimes only coarse quadrature meshes are feasible, both the asymptotic error rate \textit{and} the error prefactor need to be reduced. The interplay between these two descriptors of the FSE underpins the theoretical motivations of this work and will be emphasized throughout this section. %

\subsection{Fock Exchange SS with the Gaussian Auxiliary Function}

Following eq \eqref{eq:singularity-subtraction-general}, we define $g(\vx)=4\pi h(\vx)/|\vx|^2$, and we henceforth call $h(\vx)$ the auxiliary function. The Fock exchange energy under the SS method (using $g(\vx)$ as an argument) is then given by
\begin{align}
E_{X}^{N_{\mathbf{k}}}[g(\vq)]&=-\frac{\left|\Omega^{*}\right|}{16\pi ^3N_{\mathbf{k}}}\sum_{\mathbf{q}\in\mathcal{K}_{\mathbf{q}}}\primedsum_{\mathbf{G}\in\mathbb{L}^{*}}\left(\hat{V}_{\mathrm{coul}}(\mathbf{q}+\mathbf{G})S(\mathbf{q}+\mathbf{G})-g(\mathbf{q}+\mathbf{G})\right)\nonumber\\
&\qquad-\frac{1}{16\pi^3}\int_{\mathbb{R}^{3}}d\mathbf{q}\ g(\mathbf{q})\label{eq:exxss_g}\\
&=-\frac{\left|\Omega^{*}\right|}{16\pi ^3N_{\mathbf{k}}}\sum_{\mathbf{q}\in\mathcal{K}_{\mathbf{q}}}\primedsum_{\mathbf{G}\in\mathbb{L}^{*}}\left(S(\mathbf{q}+\mathbf{G})-h(\mathbf{q}+\mathbf{G})\right)\hat{V}_{\mathrm{coul}}(\mathbf{q}+\mathbf{G})\nonumber\\
&\qquad-\frac{1}{16\pi^3}\int_{\mathbb{R}^{3}}d\mathbf{q}\ \hat{V}_{\mathrm{coul}}(\mathbf{q})h(\mathbf{q})\label{eq:exxss_h}
\end{align}

Consider choosing a Gaussian auxiliary function, 
\begin{align}
    h(\vx) &=e^{-|\vx|^{2}/2\sigma^2}.
\end{align}
Aside from the fact that using the Gaussian auxiliary function provides a closed-form expression for the integral term\cite{massiddaHartreeFockLAPWApproach1993}, ref \citenum{holzwarthAnalysisNumericalMethods2011} also provides a justification of this auxiliary function by showing that for a simple periodic model involving non-overlapping Gaussian orbitals, the Fourier transform of the pair-densities is also Gaussian, which will lead to a Gaussian-shaped structure factor. The use of this auxiliary function thus leads to a desirable reduction in the magnitude of $S(\vq)-h(\vq)$ and thus the quadrature term.  Formally, the Gaussian auxiliary function is equivalent to applying the following $\sigma$-dependent correction to the uncorrected exchange energy, that is,%
\begin{align}
    E_X^{N_\textbf{k},\sigma}:= E_{X}^{N_{\mathbf{k}}}\left[4\pi e^{-|\vq|^{2}/2\sigma^2}/|\vq|^2\right]&=E_{X}^{N_{\mathbf{k}}}+\Nocc\xi(\sigma),\label{eq:exchange_ss1g}
\end{align}
where
\begin{align}
\label{eqn:xi_gaussian}
    \xi(\sigma) &=\left(\frac{|\Omega^*|}{2N_{\mathbf{k}}}\sum_{\mathbf{q}\in\mathcal{K}_{\mathbf{q}}}-\frac{1}{2}\int_{\Omega^{*}}d\mathbf{q}\right) \left(\frac{1}{(2\pi)^{3}}\sideset{}{'}\sum_{\mathbf{G}\in\mathbb{L}^{*}}\frac{4\pi e^{-|\mathbf{q}+\mathbf{G}|^{2}/2\sigma^2}}{|\mathbf{q}+\mathbf{G}|^{2}}\right).
\end{align}
Note that if $S(\vq)$ is a perfect Gaussian of standard deviation $\sigma$, then the correction from $\xi(\sigma)$ will be exact. %
\firstpassrev{We note that, by Theorem 5.1 in ref \citenum{xingUnifiedAnalysisFinitesize2024}, $E_X^{N_\textbf{k},\sigma}$ also converges to the TDL as $\mathcal{O}\left(\Nk^{-1}\right)$ for any positive and finite $\sigma$.} %

One of the most widely used choices is found in the Madelung constant correction \cite{fraserFinitesizeEffectsCoulomb1996,drummondFinitesizeErrorsContinuum2008,martinCoulombInteractionsExtended2004}, which is defined as
\begin{equation}\label{eqn:exchange_madelung}
    E_X^{N_\textbf{k},\text{Madelung}} = E_X^{N_\textbf{k}}+\Nocc\xi_{\text{Madelung}},
\end{equation}
where
\begin{align}
\xi_{\text{Madelung}}
&=
\left(\frac{|\Omega^*|}{2N_{\mathbf{k}}}\sum_{\mathbf{q}\in\mathcal{K}_{\mathbf{q}}}-\frac{1}{2}\int_{\Omega^{*}}d\mathbf{q}\right)
\left(\frac{1}{(2\pi)^{3}}\sideset{}{'}\sum_{\mathbf{G}\in\mathbb{L}^{*}}\frac{4\pi e^{-|\mathbf{q}+\mathbf{G}|^{2}/\eta}}{|\mathbf{q}+\mathbf{G}|^{2}}\right) \notag \\
&\quad
+\sideset{}{'}\sum_{\mathbf{R}\in\mathbb{L}_{\mathcal{K}_{\mathbf{q}}}}\frac{\operatorname{erfc}\left(\eta^{1/2}|\mathbf{R}|/2\right)}{2|\mathbf{R}|}
-\frac{2\pi}{|\Omega|N_{\mathbf{k}}\eta}.
\label{eq:madelung-regular}
\end{align}
Above, $\mathbb{L}_{\mathcal{K}_\mathbf{q}}$ is the real space lattice associated with all points $\mathbf{q}+\mathbf{G}$ generated by $\mathcal{K}_{\mathbf{q}}$ and $\mathbb{L}^{*}$ respectively. The Bravais lattice generated by $\mathbb{L}_{\mathcal{K}_\mathbf{q}}$ is exactly that generated by the supercell and $\vq+\vG$ defines the associated reciprocal lattice. Note that $\xi_{\text{Madelung}}$ is a constant for any $\eta>0$. %
Physically, the Madelung constant is an Ewald summation that computes the electrostatic interaction of a probe electron placed at the origin of the equivalent supercell in a uniform charge-compensating background and its periodic images.\cite{hubQuantifyingArtifactsEwald2014,paierPerdewBurkeErnzerhof2005}

Comparing eqs \eqref{eqn:xi_gaussian} and \eqref{eq:madelung-regular}, we find that $\xi_{\text{Madelung}}$ is equivalent to the Gaussian singularity subtraction scheme in the limit $\sigma\to \infty$, that is, $h(\vq)=1$ \cite{xingUnifiedAnalysisFinitesize2024}. As previously alluded to, if $h(\vq)$ is the exact shape of $S(\vq)$, then the SS method shown in eq \eqref{eq:singularity-subtraction-general} using the Gaussian auxiliary function will yield the exact exchange energy at the TDL. Clearly $\sigma=\infty$ is not the optimal choice of the Gaussian width. Some widely-used improvements include setting $\sigma$ to be the diameter of the FBZ\cite{massiddaHartreeFockLAPWApproach1993} and using finite differences to estimate and correct second-order errors that the Gaussian auxiliary function misses.\cite{ducheminScalableAccurateAlgorithm2010}

We may also choose $\sigma$ by directly minimizing the discrepancy between the structure factor $S(\vq+\vG)$ and the Gaussian auxiliary function $h(\vq+\vG)$. A simple way to accomplish this is by fitting a Gaussian of width $\sigma$ to $S(\vq+\vG)$ itself through least-squares regression on the reciprocal lattice points $\vq+\vG$. We will consider the points that are at most $q_\mathrm{cut}$ away from the origin,
\begin{align}
    \sigma_\mathrm{fit} &= \argmin_\sigma \primedsum_{|\vq+\vG|<q_\mathrm{cut}}\left[\frac{S(\vq+\vG)}{|\vq+\vG|^2}-\frac{N_\mathrm{occ}e^{-|\vq+\vG|^2/2\sigma^2}}{|\vq+\vG|^2}\right]^2. \label{eq:sigma_fit}
\end{align}
Note that $S(\vq)=N_\mathrm{occ}$ at the origin. In the regression, we weight via the Coulomb kernel $1/|\vq+\vG|^2$ because our objective is to find the $\sigmafit$ that yields the best estimate of $E_X^\text{TDL}$, not of the structure factor itself. In effect, this places more emphasis on the residuals evaluated at $\vq$-points closer to the origin.

\begin{figure}
    \centering
    \includegraphics[width=\textwidth]{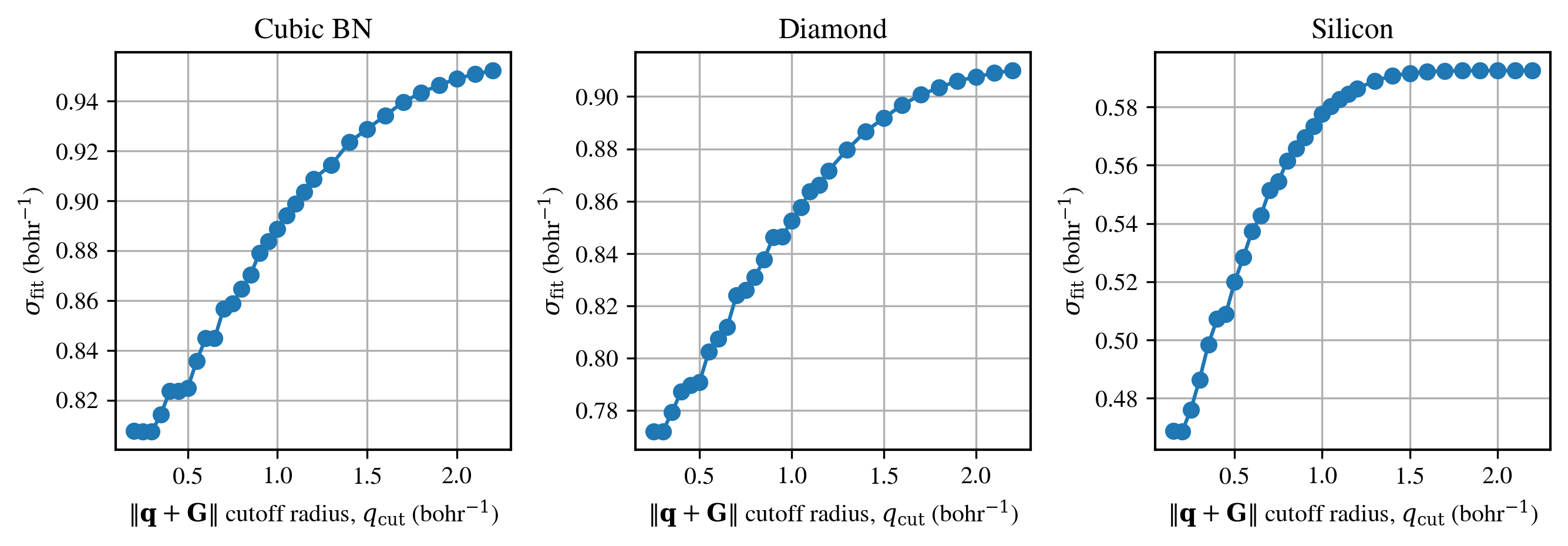}
    \caption{Fitted $\sigma$ values for Diamond, Silicon, and Cubic-BN at different $q_\mathrm{cut}$ values. Calculations were at the PBE0/SZV-GTH level of theory with $\Nk=8^3$.}
    \label{fig:sigma_vs_qcut}
\end{figure}

\begin{figure}
    \centering
    \begin{subfigure}{0.48\textwidth}
        \centering
        \includegraphics[width=\textwidth]{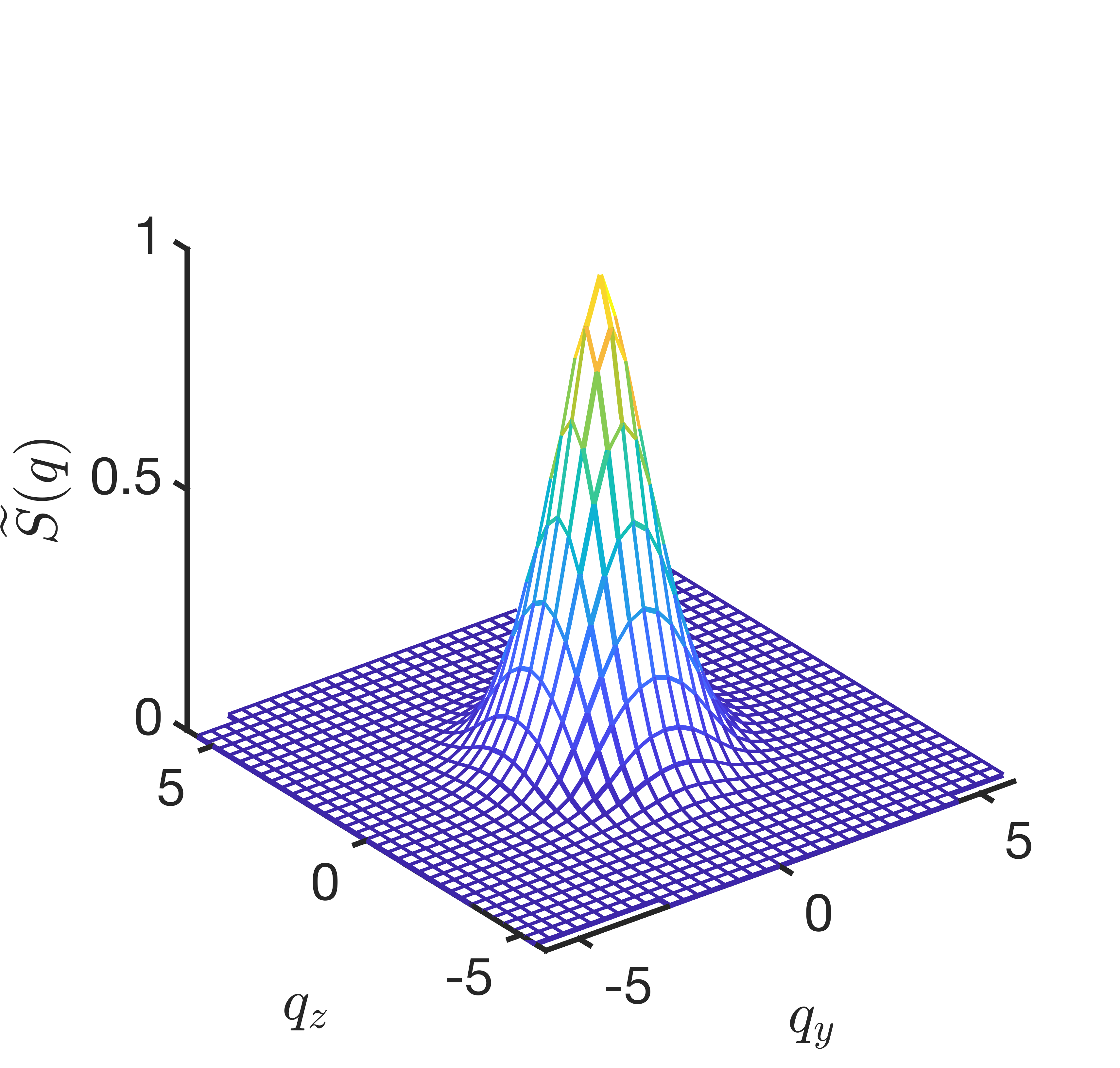}
        \caption{}
        \label{fig:diamond_SqG}
    \end{subfigure}
    \hfill
    \begin{subfigure}{0.48\textwidth}
        \centering
        \includegraphics[width=\textwidth]{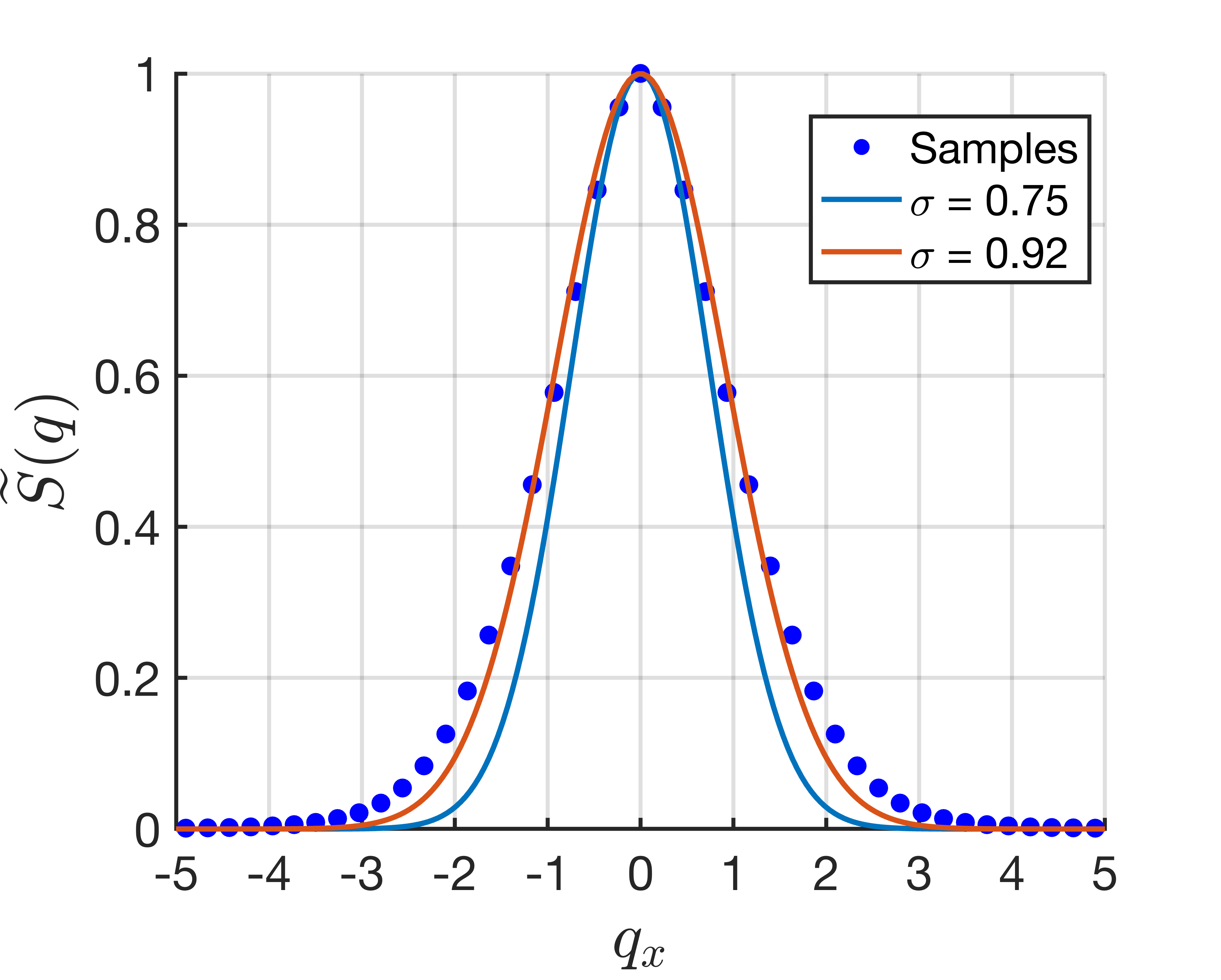} 
        \caption{}
        \label{fig:diamond_SqG_1d}
    \end{subfigure}
    \caption{Normalized structure factor, $\widetilde{S}(\vq+\vG)=S(\vq+\vG)/\Nocc$, for diamond (a) at $q_x=0$ and (b) along $q_x$, with Gaussians of width $\sigma=0.75$ and $\sigma=0.92$ bohr$^{-1}$ shown. These correspond approximately to the Gaussian widths fitted from the minimum and maximum $q_\mathrm{cut}$, respectively, as shown in Figure \ref{fig:sigma_vs_qcut}.}
    \label{fig:diamond_structure_factor}
\end{figure}

Our first observation is that, when performing this fitting procedure for the diamond, silicon, and cubic-BN crystals at $\Nk=8\times8\times8$, the fitted $\sigma$ values are sensitive to the cutoff, $q_\mathrm{cut}$, as shown in Figure \ref{fig:sigma_vs_qcut}. In fact, for these crystals, there is a systematic increase in the fitted width as the cutoff is increased.  
This is because, as shown in Figure \ref{fig:diamond_structure_factor} for the case of diamond, the structure factor tends to decay more slowly than a Gaussian, which means that as $q_\mathrm{cut}$ is increased, $\sigmafit$ will be forced to increase in order to better fit the $S(\vq+\vG)$ points at larger $q$.

\begin{figure}
    \centering
    \includegraphics[width=\textwidth]{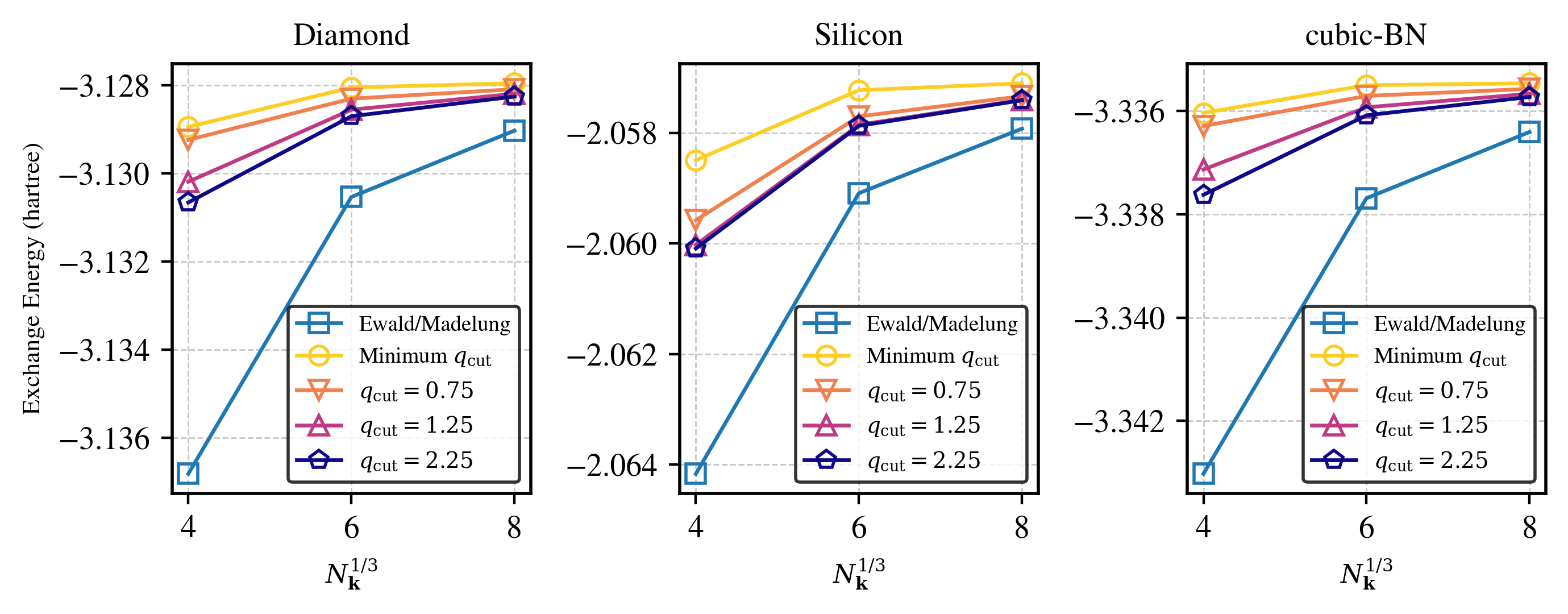}
    \caption{Finite-size error of the exchange energy for various systems at the PBE0/SZV-GTH level of theory, with the exchange energy from the Madelung correction \eqref{eqn:exchange_madelung} and from the Gaussian auxiliary function \eqref{eq:exchange_ss1g} of varying $q_\mathrm{cut}$ values shown in units of bohr$^{-1}$.} 
    \label{fig:FSE_qcut_dependence}
\end{figure}

The second observation is that better estimates of the TDL exchange energy are obtained by using smaller $q_\mathrm{cut}$ values to find $\sigmafit$ as shown in Figure \ref{fig:FSE_qcut_dependence}, and therefore the best performing Gaussian width comes from fitting to the $\vq$-points that are closest to the $\Gamma$-point.\footnote{ The fitting of a Gaussian to the points closest to the origin is similar to (but not the same as) estimating the curvature via finite differences, an approach reminiscent of the one taken in ref \citenum{ducheminScalableAccurateAlgorithm2010}. For a discussion of the comparison of these approaches, readers are directed to Section 3 of the Supplementary Information.}
This is because when the Gaussian is chosen to fit to the structure factor at the quadrature nodes further away from the origin, the second derivative of the structure factor at the origin is not as well approximated even if the structure factor as a whole is; indeed, the second derivative is what gives rise to the asymptotically leading FSE term\cite{xingUnifiedAnalysisFinitesize2024}. 

\subsection{Construction of Robust Auxiliary Functions}\label{sec:robust_aux_functions}
 
The results in the previous section reveal limitations with the use of the Gaussian auxiliary function to correct the finite-size error in the exchange energy. In Figure \ref{fig:diamond_SqG_1d}, the wider Gaussian with $\sigma=0.92$ is a better approximation to the structure factor as a whole than the narrower Gaussian with $\sigma=0.75$, yet, the narrower Gaussian yields a better estimate of the exchange energy. This trend indicates that the Gaussian auxiliary function is not optimal for simultaneously describing the behavior of the structure function near $\mathbf{q}=0$, which governs the asymptotic error scaling, and the overall form of the structure factor in real systems, which influences the error preconstant.

While the large $|\vq|$ decay of the structure factor for an incomplete Gaussian basis should be Gaussian-like as shown in ref \citenum{holzwarthAnalysisNumericalMethods2011}, the decay of the structure factor at low to moderate $|\vq+\vG|$ is evidently slower than a Gaussian. Indeed, being able to capture the behavior of the singularity close to the origin is crucial to reducing the FSE asymptotically. But if one were to also capture the derivatives of the integrand beyond the second order, we could see that the prefactor to the FSE is also reduced. This motivates the need for more flexible yet robust auxiliary functions that accurately capture the true decay behavior of the structure factor.

For inspiration, we can first exploit a general property of the Fourier transforms of smooth and periodic functions---in particular, the Fourier coefficients $\hat{f}_k$ of a smooth and periodic function decay super-polynomially or even exponentially\cite{trefethenExponentiallyConvergentTrapezoidal2014}, that is, $\hat{f}_k\sim e^{-\alpha k}$ for large enough $k$ and for some $\alpha>0$, but usually not as far as a Gaussian. This suggests that the auxiliary function should have some component that decays exponentially. We thus propose two auxiliary functions,
\begin{align}
    h_\mathrm{Exponential}(q,\{\alpha, \gamma\})&=e^{-\alpha\left(\sqrt{1+\gamma q^2}-1\right)},\\
    h_\mathrm{QuarticExponential}(q,\{\alpha, \gamma_1, \gamma_2\})&=e^{-\alpha\left(\sqrt{1+\gamma_1 q^2+\gamma_2 q^4}-1\right)},
\end{align}
where the Quartic Exponential function also has a way to account for Gaussian-like decay at large $q$ through $\gamma_2$.

An alternate approach to constructing auxiliary functions is to consider that, although the periodic parts $u_{n\mathbf{k}}(\mathbf{r})$ are indeed GTOs with phase factors and therefore result in reciprocal pair densities that would decay like a Gaussian, one of the goals of using contracted GTOs is to better describe cusps in the wavefunction near the nuclei \cite{szaboModernQuantumChemistry1982}. 
In fact, the Fourier transform of a true Slater-type orbital (STO) will exhibit \textit{algebraic} decay \cite{belkicUnifiedFormulaFourier1989}, which is much slower than the Gaussian decay rate.
Thus, the decay of the Fourier coeffficients of a contracted GTO could, for some portion of reciprocal space, decay as $1/(1+\gamma k^2)^\beta$ before exhibiting Gaussian decay at large q, with the exponent $\beta$ becoming a free parameter. This suggests the following forms of the auxiliary function
\begin{align}
    h_\mathrm{Lorentzian}(q,\{\gamma, \beta\})&=\frac{1}{(1+\gamma q^2)^\beta},\\
    h_\mathrm{GaussLorentz}(q,\{\sigma,\alpha,\beta,\gamma\})&=\left(1-e^{-\alpha q^2}\right) e^{-q^2 / 2 \sigma^2}+\left(e^{-\alpha q^2}\right) \frac{1}{\left(1+\gamma q^2\right)^\beta},\\
    h_\mathrm{GaussSlater}(q,\{\sigma,\alpha,\beta,\gamma_1,\gamma_2,\gamma_3\})&=\left(1-e^{-\alpha q^2}\right) e^{-q^2 / 2 \sigma^2}+\left(e^{-\alpha q^2}\right) \frac{1+\gamma_1 q^2+\gamma_2 q^4}{\left(1+\gamma_3 q^2\right)^\beta}.
\end{align}
where $\alpha$ in the latter two functions is a parameter that controls how fast the auxiliary function switches from Lorentzian-like decay to Gaussian decay as $q$ increases.

\begin{table}[h]
\centering
\caption{Auxiliary functions considered in this work. Labeled in asterisks are the auxiliary functions selected for testing in Section \ref{sec:results}.} 
\renewcommand{\arraystretch}{2} %
\begin{tabular}{c|c|c}
\hline
\hline
\textbf{Auxiliary Function} & \textbf{Parameters} & \textbf{Expression} \\
\hline
Gaussian* & $\sigma$ & $e^{-q^2 / (2\sigma^2)}$ \\
\hline
Exponential & $\alpha,\,\gamma$ & $e^{-\alpha\left(\sqrt{1+\gamma q^2}-1\right)}$ \\
\hline
Quartic Exponential* & $\alpha,\,\gamma_1,\,\gamma_2$ & $e^{-\alpha\left(\sqrt{1+\gamma_1 q^2+\gamma_2 q^4}-1\right)}$ \\
\hline
Lorentzian & $\gamma,\,\beta$ & $\dfrac{1}{(1+\gamma q^2)^\beta}$ \\
\hline
Gaussian-Lorentzian* & $\sigma,\,\alpha,\,\beta,\,\gamma$ & $\left[1-e^{-\alpha q^2}\right] e^{-q^2/(2\sigma^2)} + e^{-\alpha q^2} \dfrac{1}{(1+\gamma q^2)^\beta}$ \\
\hline
Gaussian-Slater & $\sigma,\,\alpha,\,\beta,\,\gamma_1,\,\gamma_2,\,\gamma_3$ & $\left[1-e^{-\alpha q^2}\right] e^{-q^2/(2\sigma^2)} + e^{-\alpha q^2} \dfrac{1+\gamma_1 q^2+\gamma_2 q^4}{(1+\gamma_3 q^2)^\beta}$ \\
\hline
\hline
\end{tabular}
\label{tab:auxiliary_functions}
\end{table}

\begin{figure}
    \centering
    \begin{subfigure}{0.48\textwidth}
        \centering
        \includegraphics[width=\textwidth]{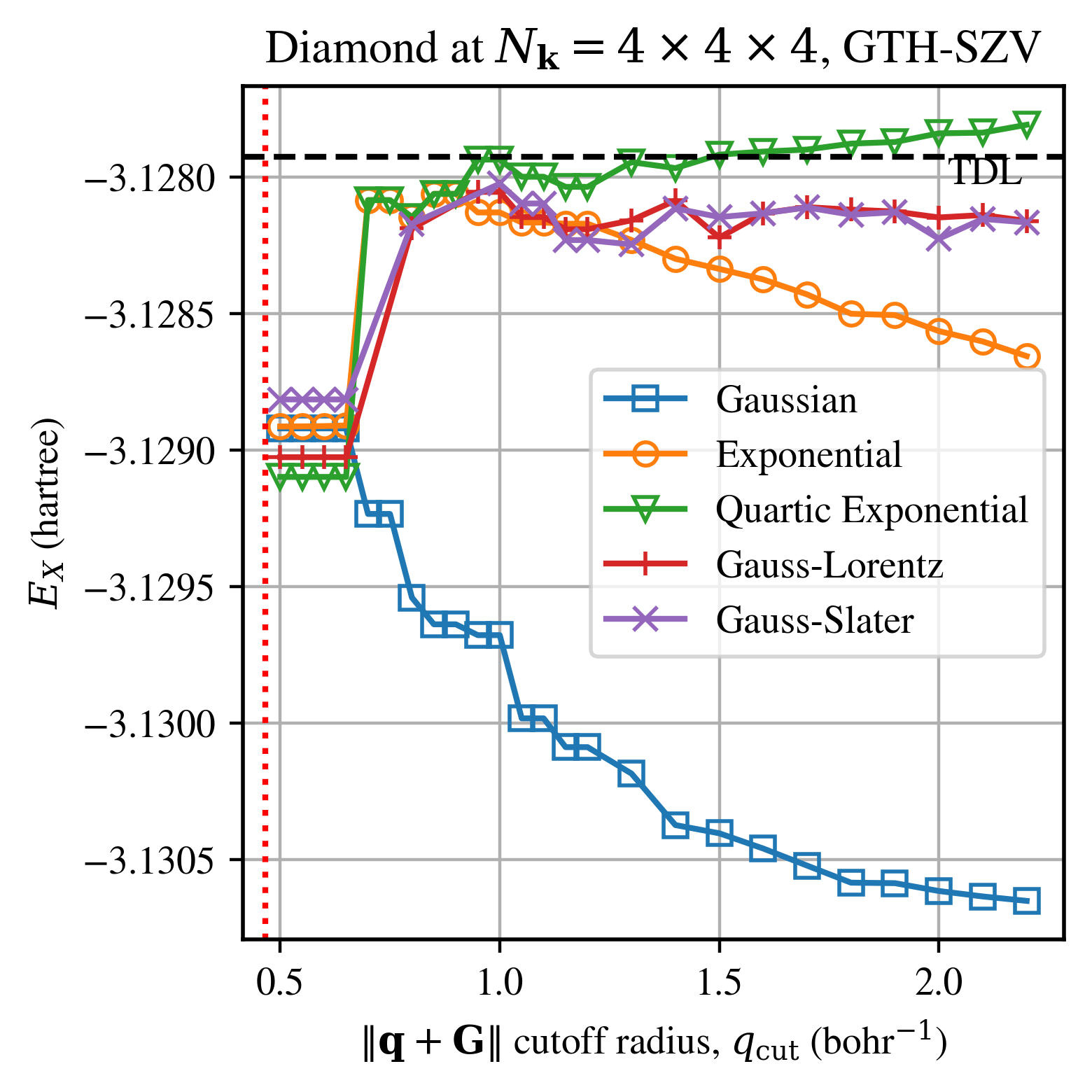}
        \label{fig:auxfunc_gthszv}
    \end{subfigure}
    \hfill
    \begin{subfigure}{0.48\textwidth}
        \centering
        \includegraphics[width=\textwidth]{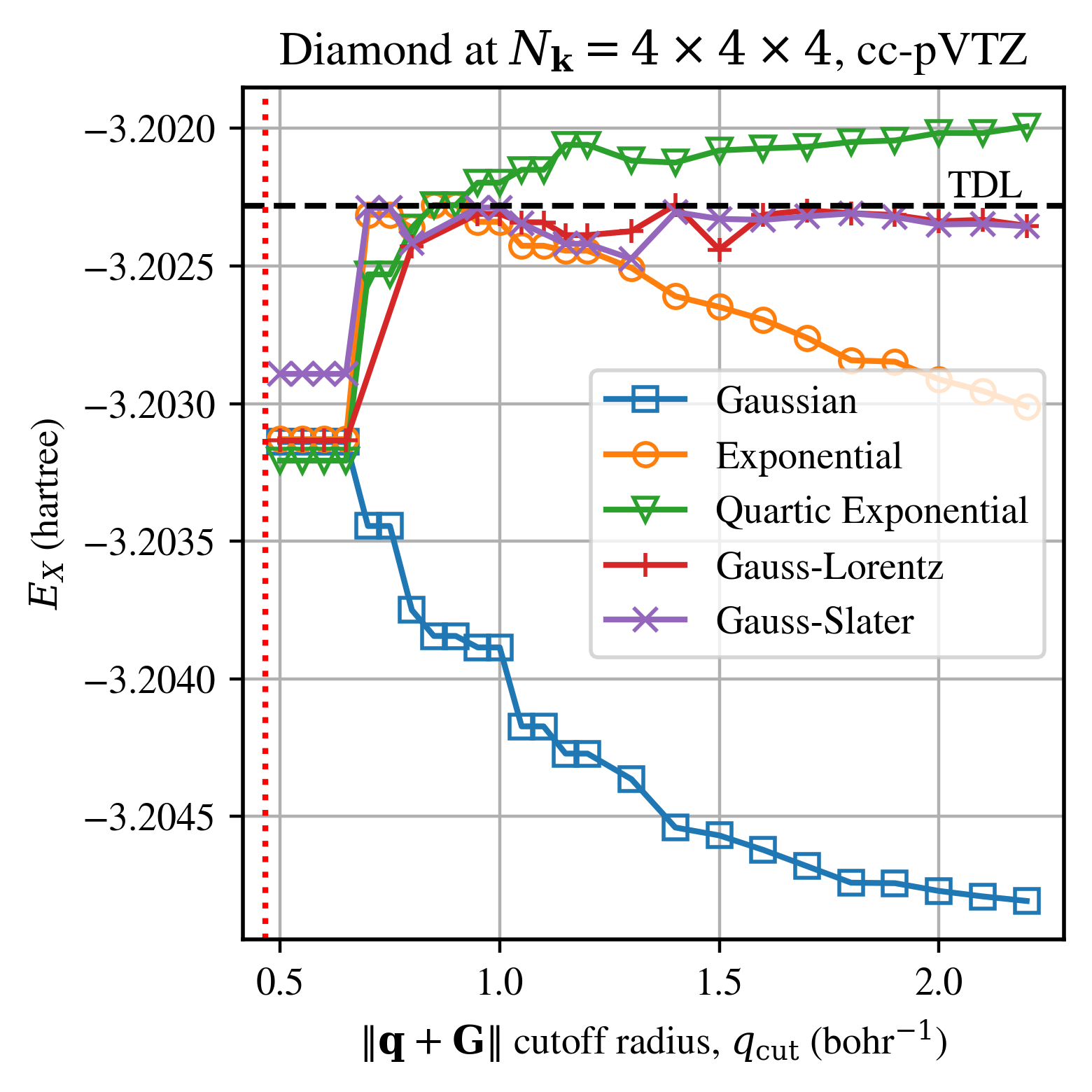}
        \label{fig:auxfunc_ccpvtz}
    \end{subfigure}
    \caption{Performance of different auxiliary functions with the PBE0 functional at different basis sets. The TDL was computed by taking the Ewald/Madelung curve and fitting an $\Nk^{-1}$ power law polynomial to the exchange energy values at $\Nk=[6^3,8^3]$ and $[4^3,6^3]$ for GTH-SZV and cc-pVTZ, respectively. The dashed vertical red line indicates the norm of the $\vq$-point closest to the origin.}
    \label{fig:auxiliary_functions_performance}
\end{figure}

All of the aforementioned auxiliary functions are collected in Table \ref{tab:auxiliary_functions}. In order to examine the performance of these auxiliary functions, we use the same least-squares fitting procedure as in eq \eqref{eq:sigma_fit} to find the optimal parameters for all auxiliary functions for a given $q_\mathrm{cut}$. We perform a PBE0 calculation on diamond with $\Nk=4\times4\times4$ and, for all auxiliary functions, compute the correction (and thus the final HF exchange energy) based on the fitted parameters as a function of $q_\mathrm{cut}$. The results are shown in Figure \ref{fig:auxiliary_functions_performance}. The Gaussian auxiliary function performs best when fitted to only the points closest to the origin. In contrast, the other auxiliary functions not only provide better estimates of the TDL value than the optimal Gaussian, but they also exhibit lower sensitivity to the choice of $q_\mathrm{cut}$. The only exception is the Lorentzian, which performs notably poorly (see Figure S3 in the Supporting Information), due to its inability to capture the exponential or Gaussian decay of the structure factor within the finite Gaussian basis set framework. Nonetheless, the results indicate that the other auxiliary functions have a degree of robustness in that they are accurately capturing the shape of $S(\vq)$ and are not solely relying on the increased flexibility from additional parameters. The Quartic Exponential, Gaussian-Lorentzian, and Gaussian-Slater auxiliary functions are clearly the best; however, it should be noted that the additional parameters incorporated by the Gaussian-Slater function appear to provide little benefit over the Gaussian-Lorentzian auxiliary function, as the curves are essentially on top of each other.

Figure \ref{fig:auxiliary_functions_performance} also shows that with a larger basis set, cc-pVTZ, the Gaussian-Lorentzian and Gaussian-Slater auxiliary functions perform more robustly. This gives credence to the idea that, as the atom-centered cusps become more pronounced with larger basis sets, auxiliary functions that include some portion of algebraic decay could be slightly more effective than an exponentially decaying one. However, we note that the Exponential and (especially) the Quartic Exponential auxiliary functions are still excellent choices, deviating from the TDL by at most a millihartree and again performing much better than the Gaussian auxiliary function.

\begin{figure}[H]
    \centering
    \begin{subfigure}{0.48\textwidth}
        \centering
        \includegraphics[width=\textwidth]{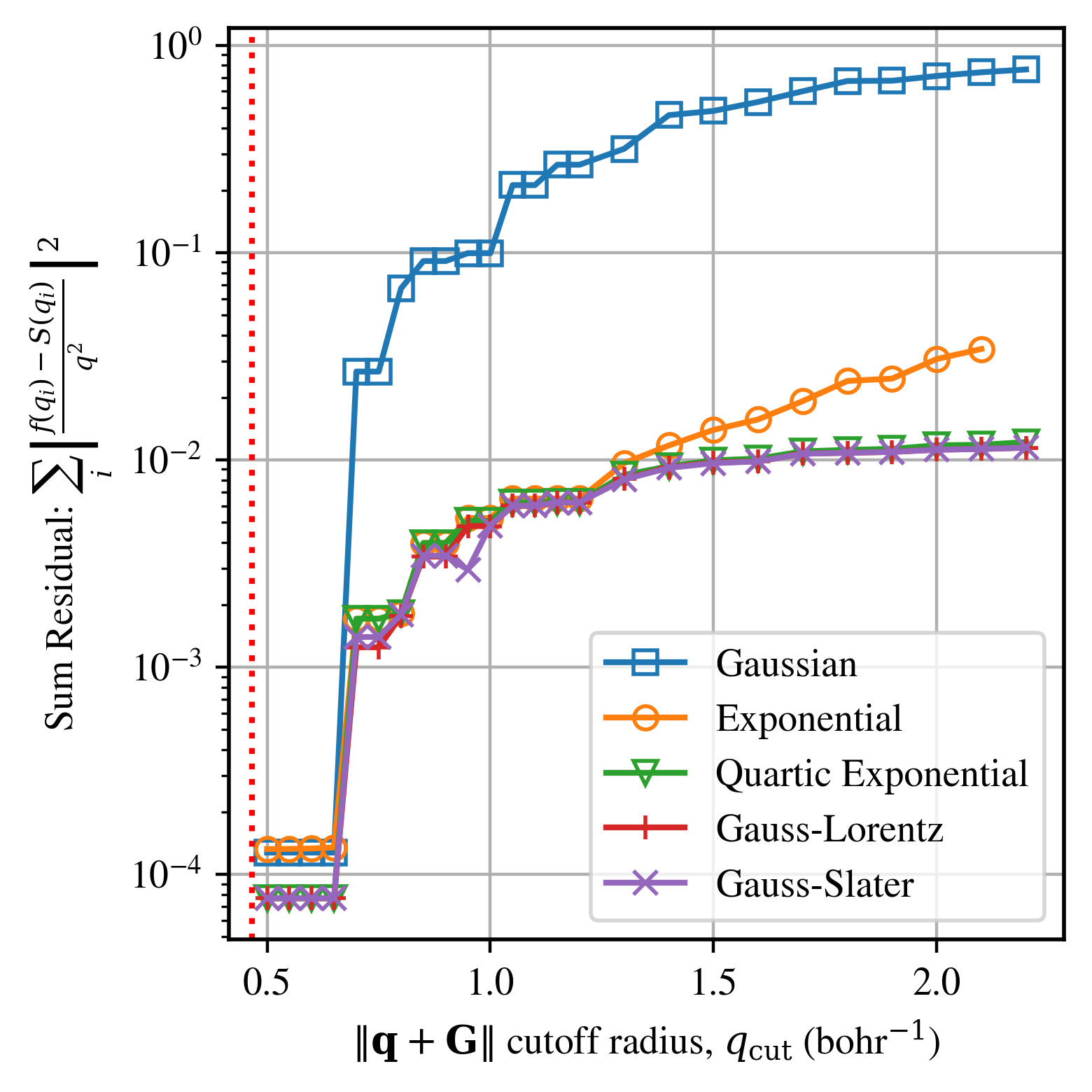}
        \caption{Auxiliary function residuals}
        \label{fig:auxfunc_residuals}
    \end{subfigure}
    \hfill
    \begin{subfigure}{0.48\textwidth}
        \centering
        \includegraphics[width=\textwidth]{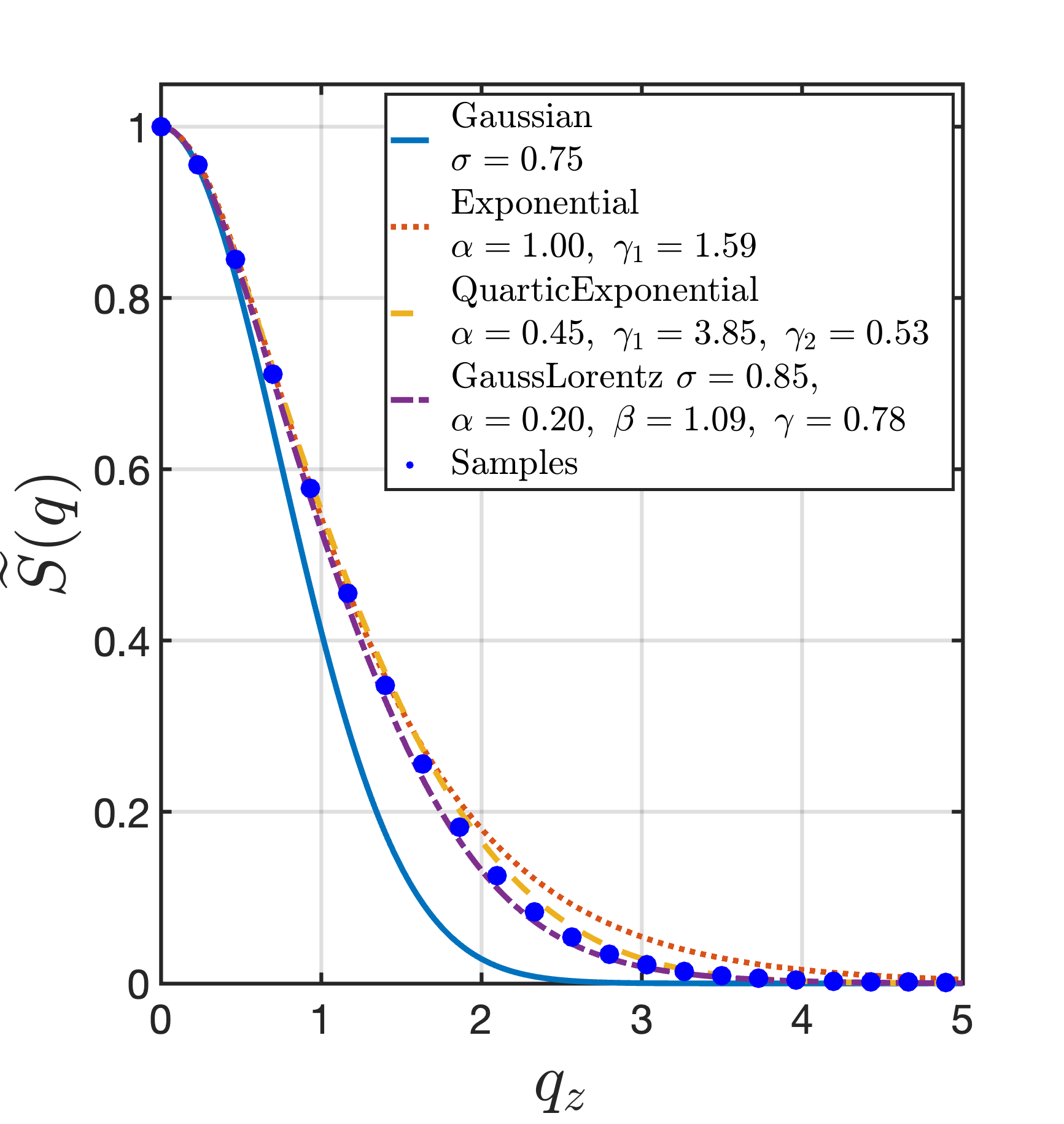}
        \caption{Visualization of fits}
        \label{fig:auxfunc_diamond_qx}
    \end{subfigure}
    \caption{Fit quality of different auxiliary functions for diamond with PBE0/GTH-SZV at $\Nk=4\times4\times4$}
    \label{fig:mainfig}
\end{figure}

We also examine the quality of the auxiliary function fits directly through the residuals and how well they capture the structure factor in Figures \ref{fig:auxfunc_residuals} and \ref{fig:auxfunc_diamond_qx}, respectively. These figures show that while capturing the behavior of the singularity around the origin is most important for removing the leading order terms in the finite size error arising from the deviation in the curvature, being able to capture some behavior of the decay at moderate to large $|\vq+\vG|$ could significantly reduce the FSE.

Therefore, in addition to the conventional Ewald/Madelung correction and the fitted Gaussian auxiliary function, we select two auxiliary functions to test their performance in reducing the exchange energy FSE: the Gaussian-Lorentzian and the Quartic Exponential. These functions, representative of the types of structure factor decay they are able to capture, were chosen based on their promising performance as shown in Figure \ref{fig:auxiliary_functions_performance} while incorporating additional parameters only when they have been shown to meaningfully reduce the FSE.

\subsection{Computational Procedure for ExxSS}\label{subsec:exxss_procedure}
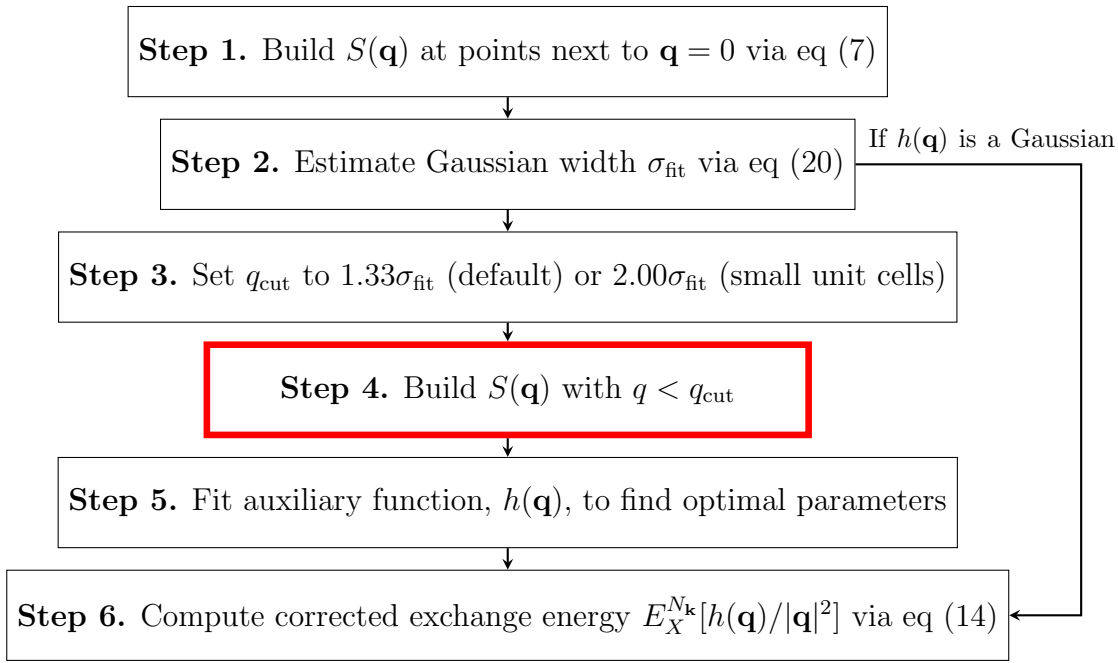
\begin{figure}[ht]
    \centering

    \begin{tikzpicture}[node distance=1.5cm]
    
    \node (step1) [process] {\textbf{Step 1.} Build $S(\mathbf{q})$ at points next to $\mathbf{q} = 0$ via eq \eqref{eq:structure_factor}};
    \node (step2) [process, below of=step1] {\textbf{Step 2.} Estimate Gaussian width $\sigmafit$ via eq \eqref{eq:sigma_fit}};
    \node (step3) [process, below of=step2] {\textbf{Step 3.} Set $q_{\text{cut}}$ to $1.33\sigmafit$ (default) or $2.00\sigmafit$ (small unit cells)} ;
    \node (step4) [highlight, below of=step3] {\textbf{Step 4.} Build $S(\mathbf{q})$ with $q < q_{\text{cut}}$};
    \node (step5) [process, below of=step4] {\textbf{Step 5.} Fit auxiliary function, $h(\vq)$, to find optimal parameters};
    \node (step6) [process, below of=step5] {\textbf{Step 6.} Compute corrected exchange energy $E_{X}^{N_{\mathbf{k}}}[h(\vq)/|\vq|^2]$ via eq \eqref{eq:exxss_h}};
    
    \draw [arrow] (step1) -- (step2);
    \draw [arrow] (step2) -- (step3);
    \draw [arrow] (step3) -- (step4);
    \draw [arrow] (step4) -- (step5);
    \draw [arrow] (step5) -- (step6);
    
    \coordinate[right=3cm of step2] (temp1);
    \coordinate[below=6cm of temp1] (temp2);
    \draw [arrow] (step2.east) -- (temp1) -- (temp2) -- (step6.east);
    
    \node[above left=0.0cm and -0.6cm of temp1, text width=3.3cm, align=right] {\footnotesize If $h(\vq)$ is a Gaussian};
    
    \end{tikzpicture}

    \caption{Flowchart of steps to compute the ExxSS exchange energy correction after the SCF routine is complete. Highlighted in red is the walltime bottleneck. See text for discussion on walltime scaling and on the recommendation for $q_\mathrm{cut}$.}
    \label{fig:sqg_flowchart}
\end{figure}

For a given auxiliary function $h(\vq)$ with its associated parameters, our objective is to compute the corrected exchange energy $E_{X}^{N_{\mathbf{k}}}\left[4\pi h(\vq)/|\vq|^2\right]$ via eq \eqref{eq:exxss_h} after the normal SCF procedure is complete. We use the following procedure for the ExxSS framework in Figure \ref{fig:sqg_flowchart}. The latter integral term in eq \eqref{eq:exxss_h} can be written as a non-singular one-dimensional integral over $q$ because the bare Coulomb kernel cancels out with the spherical coordinate Jacobian. Apart from the Gaussian auxiliary function, this integral term is computed with SciPy's adaptive quadrature\cite{piessensQuadpack1983}, reporting an error of at most $10^{-8}$ for all cases. 

The cost of the ExxSS routine depends on the chosen auxiliary function. If $h(\vq)$ is a Gaussian, then the structure factor is built only once at points next to the $\Gamma$-point to compute $\sigmafit$, resulting in an overall scaling of $\mathcal{O}\left(\Nk\right)$. For other auxiliary functions, we perform the aforementioned step and then set $q_\mathrm{cut}$ to a factor of $\sigmafit$ 
and then build the structure factor a second time for all points less than $q_\mathrm{cut}$ away from the origin. This second building scales as $\mathcal{O}\left(\Nk^2\right)$ and comprises the computational bottleneck of the ExxSS procedure. 

For the non-Gaussian auxiliary functions, we can determine an appropriate $q_\mathrm{cut}$ based on $\sigmafit$ for the given system. We revisit the results for diamond in Figure \ref{fig:auxiliary_functions_performance} which show that when $q_\mathrm{cut}$ is set to approximately 1.0 bohr$^{-1}$ (or 1.33 times the best estimate of diamond's $\sigmafit$ value of $0.75$ bohr$^{-1}$), the auxiliary functions no longer experience overfitting, and their optimal parameters may be reliably determined. We therefore recommend $q_\mathrm{cut}=1.33\sigmafit$ as the default value for ExxSS with a non-Gaussian auxiliary function. Testing indicates that for smaller unit cells containing 2-4 atoms, larger factors of $\sigmafit$ may be used to gain even more accurate estimates of the optimal parameters without incurring significant additional computational cost. However, as shown in Figure \ref{fig:auxiliary_functions_performance}, setting $q_\mathrm{cut}$ beyond the recommended value of $1.33\sigmafit$ is expected to result in at most an order of $10^{-4}$ hartree difference in the Fock exchange energy.

\begin{figure}
    \centering
    \includegraphics[width=\textwidth]{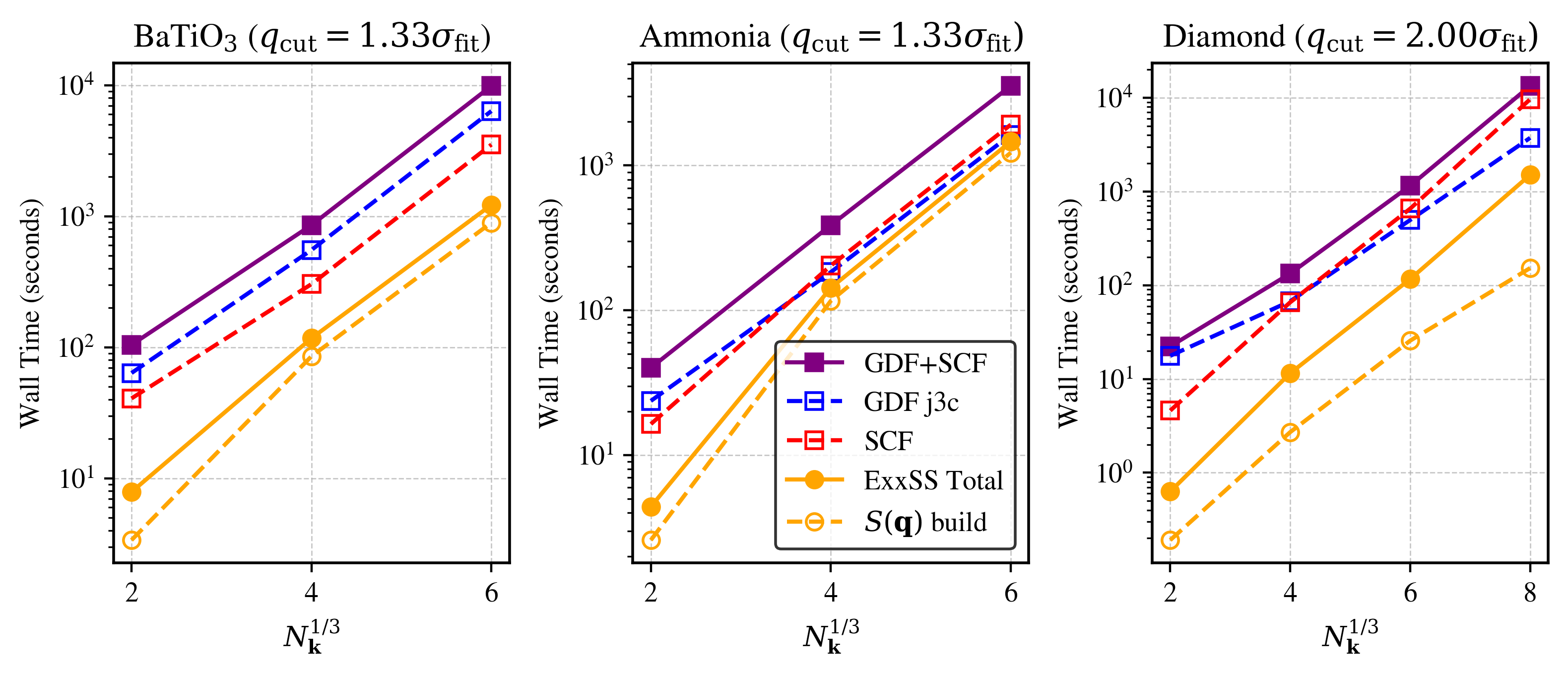}
    \caption{Walltime benchmarks for ExxSS (using a non-Gaussian auxiliary function) compared to SCF and GDF Cholesky vector building for \ce{BaTiO3}, the ammonia crystal, and diamond. The original kinetic energy cutoffs of 100, 56, and 56 hartree, respectively, were reduced to 50, 30, and 30 hartrees for the building of the structure factor. The value of $q_\mathrm{cut}$ was set to $1.33\sigmafit$ except for diamond, whose $q_\mathrm{cut}$ was set to $2.00\sigmafit$.}
    \label{fig:all_walltime_benchmarks}
\end{figure}

Furthermore, we can reduce the walltime of step 4 in Figure \ref{fig:sqg_flowchart} by evaluating the cell-periodic portion $u_{n\mathbf{k}}(\mathbf{r})$ on a coarser real-space grid (subsequently leading to fewer $\vq,\vG$ to evaluate the structure factor on). 
This is because we are building the structure factor only to obtain optimal parameters for the auxiliary function, not to compute the exchange energy directly. This change only resulted in at most approximately $10^{-5}$ hartree in the Fock exchange energy, and all figures presented in the results section were performed with the structure factor built at a coarse grid density corresponding to half of the original kinetic energy cutoff. In Figure \ref{fig:all_walltime_benchmarks}, we demonstrate that when using a coarser grid for the structure factor, the walltime of the ExxSS method is only a fraction of the SCF procedure and the building of the GDF Cholesky vectors. These benchmarks were performed on an Intel Xeon Skylake 6130 CPU using 2 threads (using more threads resulted in the SCF part of walltime \textit{increasing}).

\section{Results}\label{sec:results}
We now demonstrate how the ExxSS treatment of the exact exchange energy reduces finite-size errors, thereby improving the accuracy of computed energies and properties. Specifically, we will substitute the 25\%-scaled exact exchange component of the PBE0 functional\cite{perdewGeneralizedGradientApproximation1996,ernzerhofAssessmentPerdewBurke1999,adamoReliableDensityFunctional1999} with 25\% of the Fock exchange value provided by ExxSS. Unless noted otherwise, we use the small GTH-SZV basis set\cite{vandevondeleQuickstepFastAccurate2005} to generate dense $\vk$-point mesh values, with the exception of the \ce{BaTiO3} for which we use GTH-SZV-MOLOPT-SR\cite{vandevondeleQuickstepFastAccurate2005}. As a test of transferability, we also include results obtained with larger basis sets, GTH-DZVP and GTH-cc-pVTZ\cite{yeCorrelationConsistentGaussianBasis2022a}. The ExxSS method has been implemented in a development version of PySCF \cite{sunRecentDevelopmentsPySCF2020a}. All geometries (except ammonia) were obtained from the Materials Project\cite{jainHighthroughputInfrastructureDensity2011}, which were optimized using the PBE functional \cite{perdewGeneralizedGradientApproximation1996}; the ammonia crystal geometry was obtained from the X23 dataset \cite{dolgonosRevisedValuesX232019}.  For computing pair densities in the two-electron integrals, we use PySCF's Gaussian Density Fitting (GDF)\cite{sunGaussianPlanewaveMixed2017a}, although we note that the finite-size error does not depend on the choice of the density fitting. To avoid high-symmetry points in the Brillouin zone, we use only even-numbered Monkhorst-Pack (MP) grids, which are shifted by half a grid spacing in each direction to sidestep the $\Gamma$-point. For lonsdaleite (also called hexagonal diamond), the number of $\vk$-points along the $z$-direction was halved due to $z$-lattice vector being larger than the $x$- and $y$-lattice vectors. Finally, the value of $q_\mathrm{cut}$ for all auxiliary functions other than the Gaussian was set to $2.00\sigmafit$, which is larger than the default value of $1.33\sigmafit$ but still resulted in fast build times for the structure factor. The exceptions for $q_\mathrm{cut}$ are in the largest unit cells in our dataset---ammonia crystal and $\ce{BaTiO3}$---for which we set $q_\mathrm{cut}$ exactly to $1.33\sigmafit$. %
 
\paragraph{Fock Exchange Energies} 

\begin{figure}[H]
    \centering
    \includegraphics[width=\textwidth]{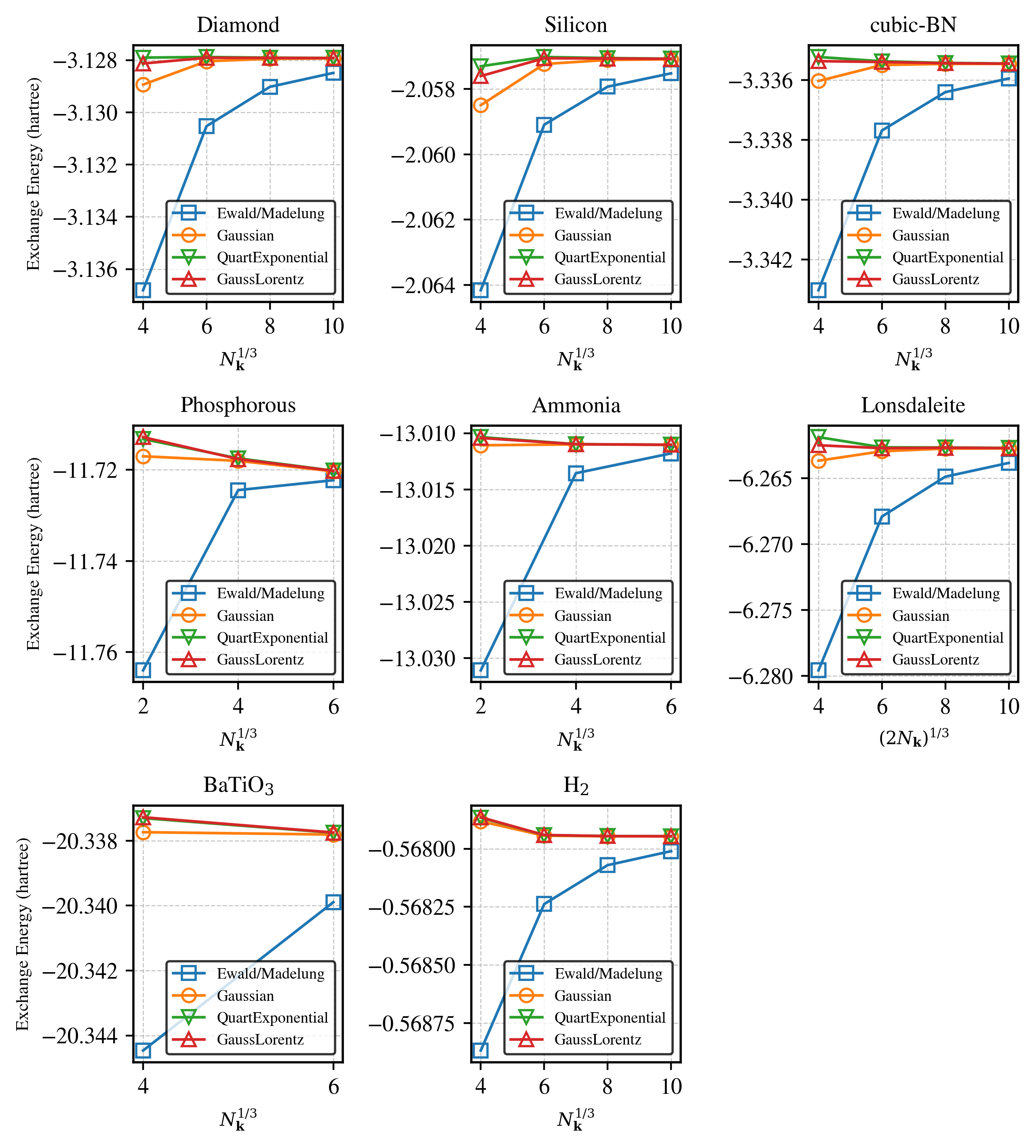}
    \caption{Finite-size effect of the exact exchange energy (before 25\% scaling) of PBE0 for various systems with the GTH-SZV basis.}
    \label{fig:all_exchange_fse}
\end{figure}

Using the density from the converged PBE0/GTH-SZV calculations, we compute (1) the Madelung-corrected exact exchange energy (before it is scaled by 0.25 in PBE0, and not to be confused with the semilocal exchange functional from PBE) and (2) the ExxSS exchange energy using the aforementioned auxiliary functions and track their convergence to the TDL as the number of $\vk$-points increases. The results, shown in Figure \ref{fig:all_exchange_fse}, demonstrate that ExxSS with the three auxiliary functions considered here outperforms the Madelung correction, achieving millihartree (and oftentimes better) accuracy toward the TDL at smaller MP $\vk$-point meshes across a variety of crystals. Notable is ExxSS's performance for the ammonia crystal, phosphorus, and the perovskite \ce{BaTiO3}, whose large unit cells restrict the use of larger $\vk$-point meshes. However, the larger unit cells result in reciprocal lattices that are more finely sampled, allowing the three tested auxiliary functions to still accurately estimate the TDL even at low $\Nk$, especially compared to the conventional Madelung correction.

\begin{figure}[H]
    \centering
    \includegraphics[width=0.9\textwidth]{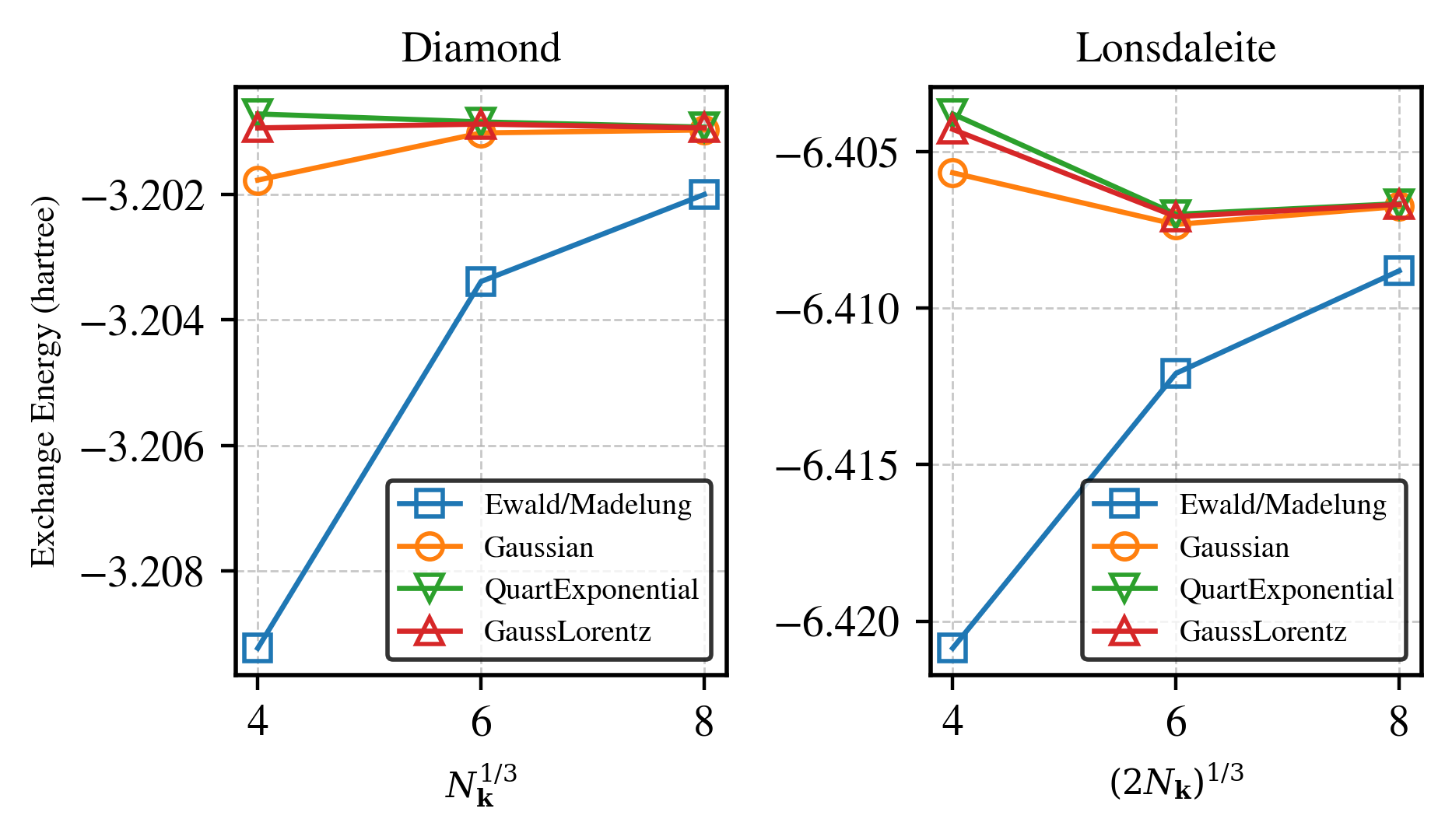}
    \caption{Finite-size effect of the exact exchange energy portion (before 25\% scaling) of PBE0 for diamond and lonsdaleite with the GTH-DZVP basis set.}
    \label{fig:all_exchange_fse_ccpvdz}
\end{figure}

\begin{figure}[H]
    \centering
    \includegraphics[width=0.9\textwidth]{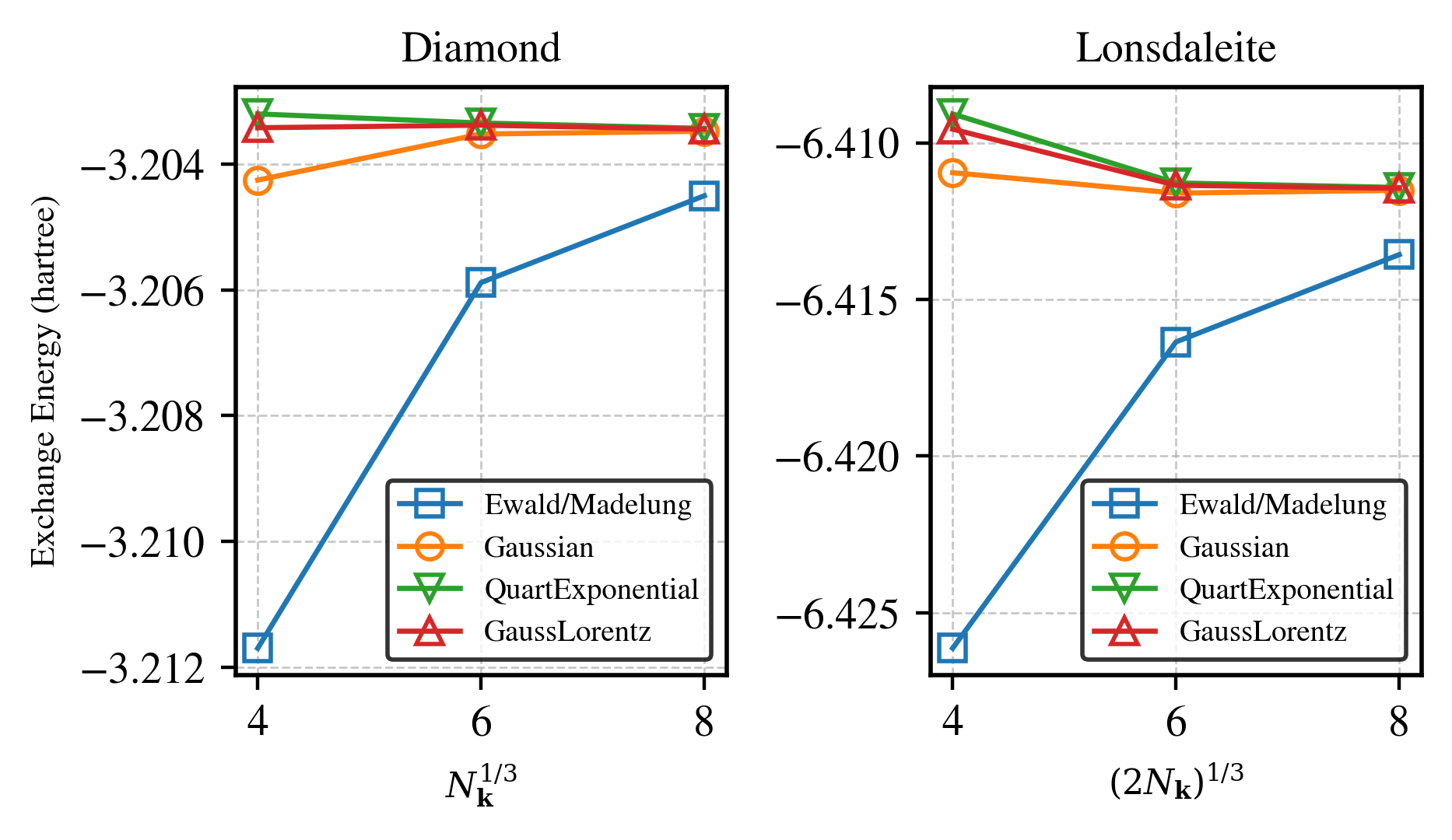}
    \caption{Finite-size effect of the exact exchange energy portion (before 25\% scaling) of PBE0 for diamond and lonsdaleite with the GTH-cc-pVTZ basis set.}
    \label{fig:all_exchange_fse_gthccpvtz}
\end{figure}

We also test the performance of the ExxSS method with the GTH-DZVP and the GTH-cc-pVTZ basis, which are the largest basis sets with which we could reliably compute $\Nk=8\times8\times8$ and $\Nk=8\times8\times4$ values for diamond and lonsdaleite, respectively. In order to overcome the disk-space bottleneck, we used the Gaussian and Planewaves density fitting scheme, also called Fast Fourier Transform Density Fitting (FFTDF)\cite{lippertHybridGaussianPlane1997,vandevondeleGaussianBasisSets2007} instead of GDF. The results are shown in Figures \ref{fig:all_exchange_fse_ccpvdz} and \ref{fig:all_exchange_fse_gthccpvtz}, once again showing that the ExxSS method with any of the auxiliary functions is able to achieve millihartree accuracy toward the TDL at smaller MP $\vk$-point meshes than the Madelung correction. We note, however, for lonsdaleite that at $\Nk=4\times4\times2$, the corrected exchange energy for the tested auxiliary functions appears to be above the TDL in both basis sets.  This is most likely because the density has not yet fully converged with respect to the $\vk$-mesh size, as
the use of a $\Gamma$-centered grid for the $\vk$-point mesh shifts the $4\times4\times2$ point below the TDL (see Figure S4 in the Supporting information). We nonetheless conclude that once the density has converged with respect to $\Nk$, the fitted auxiliary functions tested here are by far the best option for obtaining the TDL value. 

Furthermore, we highlight that at these larger basis sets, the Gaussian-Lorentzian auxiliary function appears to be the best option, even achieving virtually instant convergence in the case of diamond. This is consistent with the finding in Section \ref{sec:robust_aux_functions} that auxiliary functions that include some component of algebraic decay could see better results than ones fully relying on exponential decay. We hasten to add, however, that the quartic exponential still provides exchange energies that are very close to those provided by Gaussian-Lorentzian, all while having one less parameter. Therefore, the use of the quartic exponential function is still appropriate in this context.

\begin{figure}[H]
    \centering
    \includegraphics[width=0.7\textwidth]{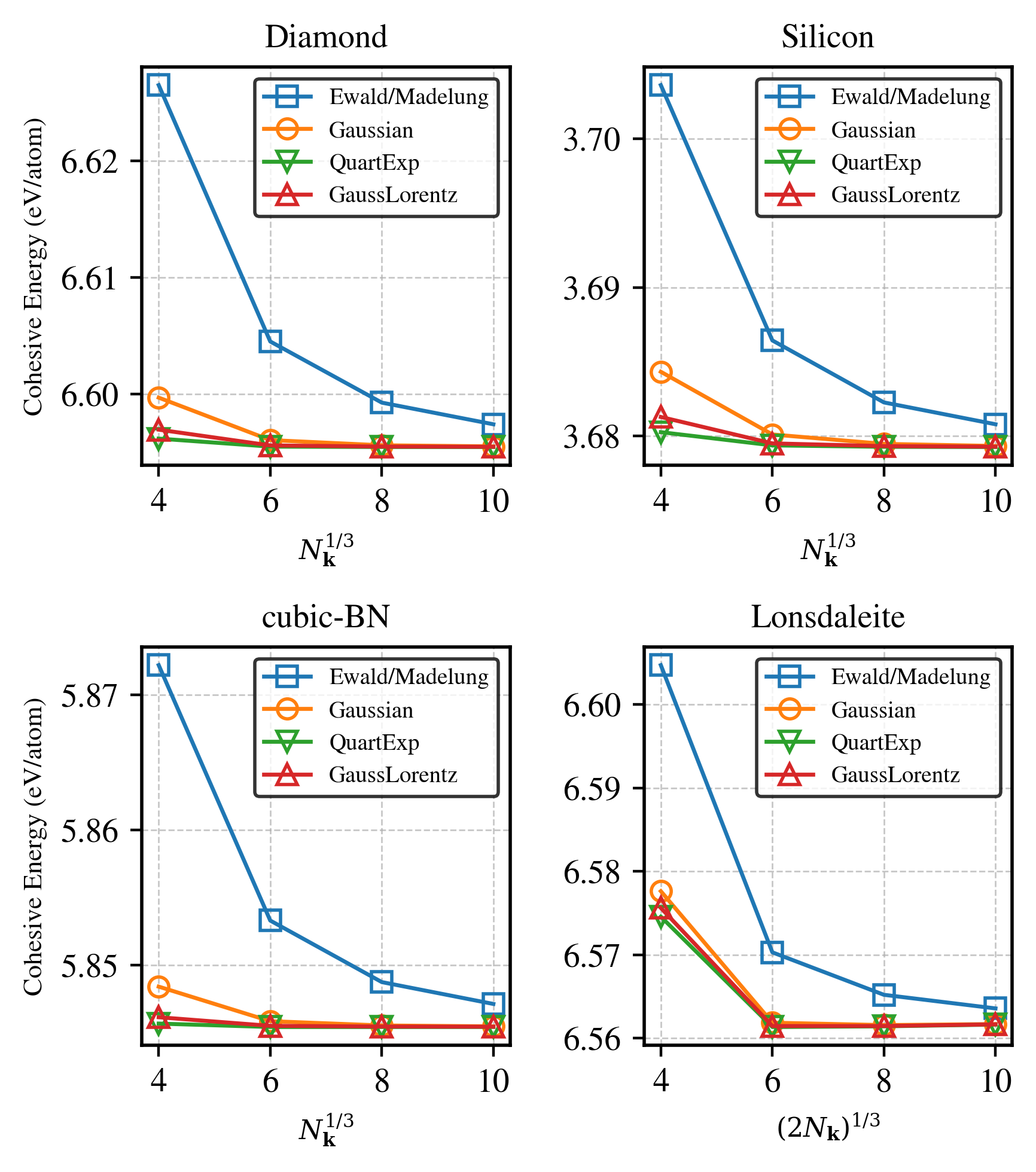}
    \caption{Finite-size effect of PBE0/GTH-SZV cohesive energies for various systems. }
    \label{fig:all_cohesive_fse}
\end{figure}

\paragraph{Cohesive Energies}
We compute the cohesive energy, defined as $E_\mathrm{coh}=\sum{E(\mathrm{atom})}-E(\mathrm{crystal})$ and computed on a per-atom basis. For each type of atom, we compute $E(\mathrm{atom})$ by performing an unrestricted DFT calculation at the same level of theory (PBE0/GTH-SZV) and placing the atom in a large unit cell (100 bohr cube). \firstpassrev{For each atom type, we perform unrestricted DFT using appropriate ground state spin multiplicity (i.e., S=5 for C and Si, S=7 for N, and S=3 for B).} The improvement from ExxSS with the indicated auxiliary functions is shown in Figure \ref{fig:all_cohesive_fse}, where energies are displayed in eV/atom. To validate these calculations, we also confirm that the values computed are within reasonable deviation from experiment---7.39 eV/atom for diamond\cite{fahyVariationalQuantumMonte1990}, 4.63 eV/atom for silicon\cite{kittelIntroductionSolidState2005}, 6.60 eV/atom for c-BN \cite{lamHighDensityPhases1990}. While experimental lonsdaleite cohesive energies are difficult to find, prior \textit{ab initio} calculations show that its cohesive energy is only about 50 meV/atom different from diamond\cite{wangInitioElasticConstants2003,goelCommentIncipientPlasticity2018}, indicating that its experimental value should also not be far from that of diamond. Any remaining error from the experimental values can still come from basis set incompleteness and a failure to fully capture electron correlation, which are not the primary focus of this work.

\begin{figure}
    \centering
    \includegraphics[width=\textwidth]{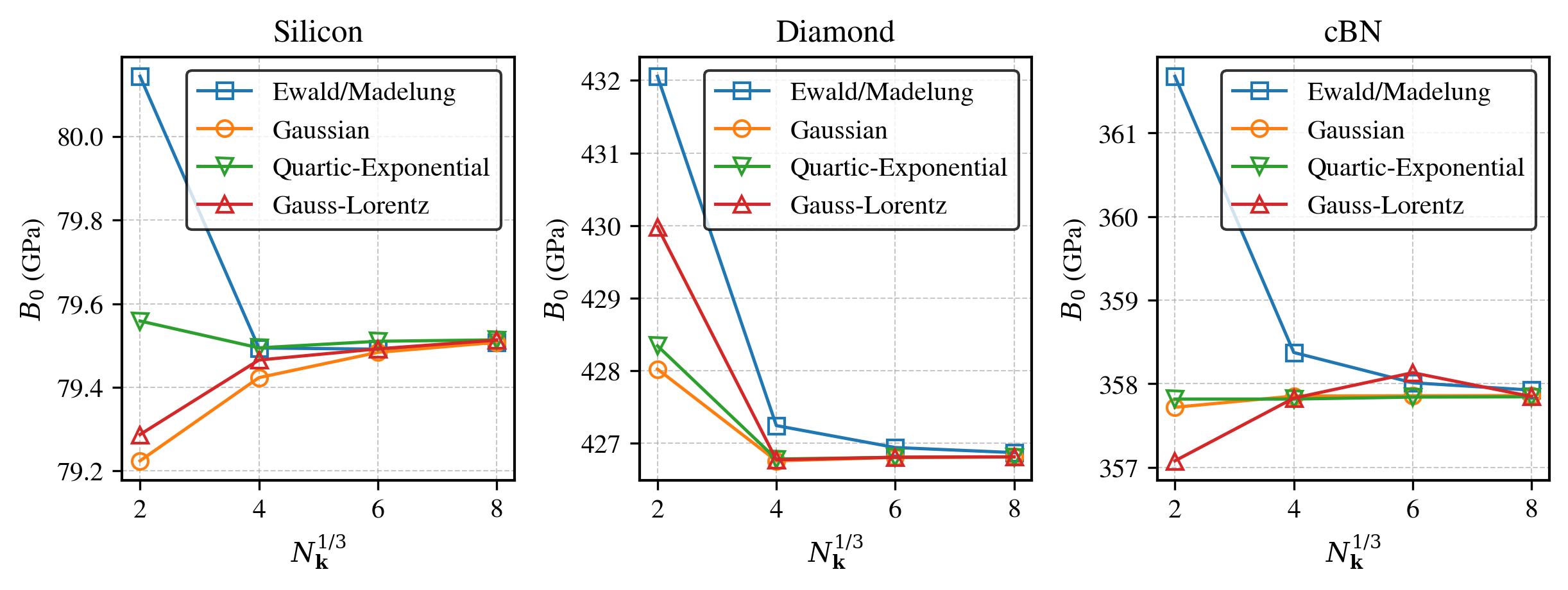}
    \caption{Finite-size effect of PBE0/GTH-SZV bulk moduli for various systems.}
    \label{fig:all_bulk_moduli_fse}
\end{figure}

\begin{figure}
    \centering
    \includegraphics[width=\textwidth]{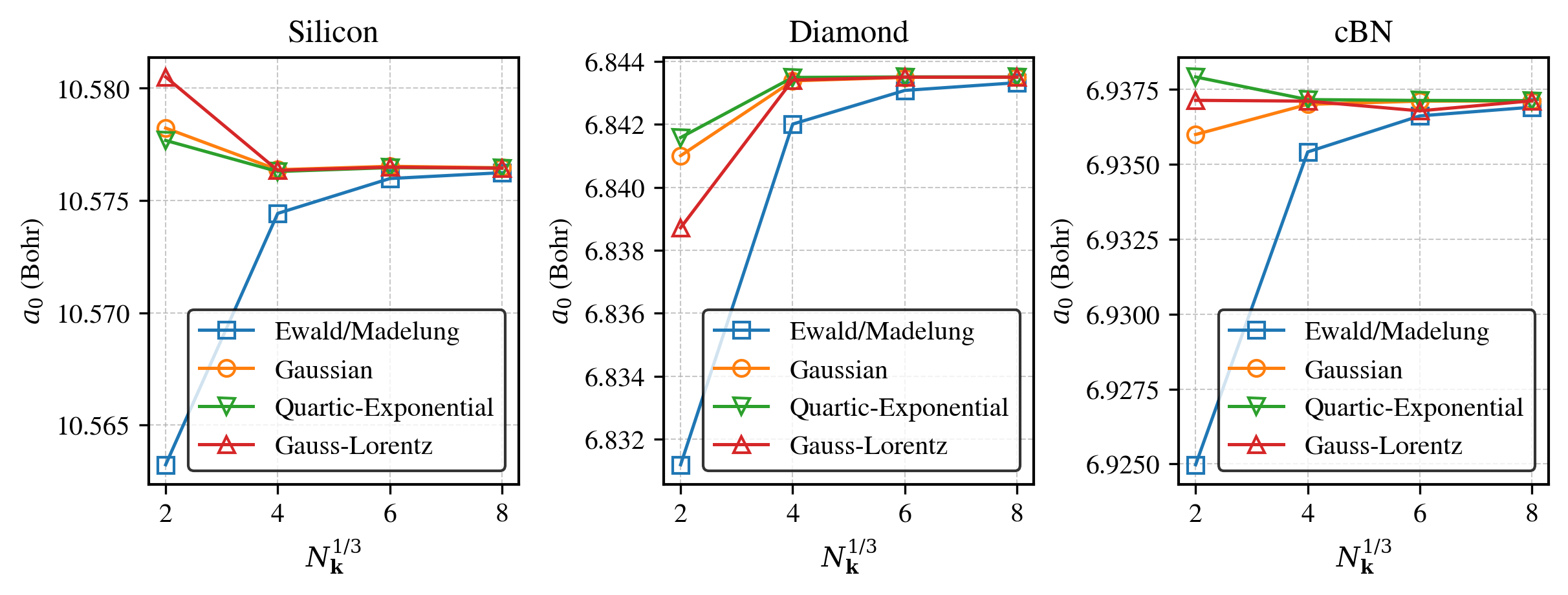}
    \caption{Finite-size effect of PBE0/GTH-SZV equilibrium lattice parameter for various systems.}
    \label{fig:all_lattice_parameter_fse}
\end{figure}

\paragraph{Bulk Modulus and Equilibrium Lattice Parameter}
The FSE also affects properties that come from derivatives of the energy, including the bulk modulus $B_0$ and the equilibrium lattice parameter $a_0$. To compute these values, we track the ground-state energy of the crystal as a function of the lattice parameter, and perform a least-squares fit of the data to the Birch-Murnaghan equation of state\cite{birchFiniteElasticStrain1947,shollDensityFunctionalTheory2009}, provided below:
\begin{align}
E_{\mathrm{total}}(a)= E_0+\frac{9 V_0 B_0}{16}\left\{\left[\left(\frac{a_0}{a}\right)^2-1\right]^3 B_0^{\prime}\right. \left.+\left[\left(\frac{a_0}{a}\right)^2-1\right]^2\left[6-4\left(\frac{a_0}{a}\right)^2\right]\right\}. \label{eq:birch-murnaghan}
\end{align}
Alongside $B_0$, and $a_0$, we are also able to extract the equilibrium total energy, $E_0$, and the bulk modulus derivative, $B'_0=(\partial B/\partial P)_T$, as parameters. The results for $B_0$ and $a_0$ are shown in figures \ref{fig:all_bulk_moduli_fse} and \ref{fig:all_lattice_parameter_fse} respectively, which show consistently the superiority of ExxSS with any of the highlighted auxiliary functions over the conventional Madelung constant and highlights how influential the FSE can be in introducing errors in these properties---up to 5 GPa in $B_0$ and 0.01 bohr in $a_0$. We note for cubic-BN that there is a slight loss in performance for the Gaussian-Lorentzian auxiliary function at $\Nk=6\times6\times6$. This is likely due to the higher number of parameters, which could result in the least squares fitting procedure finding slightly different optimal parameters as the unit cell lattice parameter changes. The Quartic Exponential, having one less parameter, does not appear to suffer the same issues in any of our tests.

Once more, we can compare the results with experimental bulk properties for validity. For $B_0$, we have experimental values of 443 GPa for diamond\cite{donohueStructuresElements1982}, 369 GPa for c-BN\cite{zaouiStructuralElectronicProperties1999,devriesCubicBoronNitride1972}, 98.0 GPa for silicon\cite{huCrystalDataHighpressure1986}. For experimental $a_0$, we have 6.741 bohr for diamond\cite{mcskiminElasticStiffnessModuli1972}, 6.833 bohr for c-BN\cite{somaCharacterizationWurtziteType1974}, and 10.27 bohr for silicon\cite{huCrystalDataHighpressure1986} respectively. The results in Figures \ref{fig:all_bulk_moduli_fse} and \ref{fig:all_lattice_parameter_fse} are consistent with PBE0 slightly underestimating $B_0$ and overestimating $a_0$ for all systems in a previous benchmark by Tran and coworkers\cite{tranRungs142016}.

\section{Discussion}

By addressing the slow convergence of the exact exchange component in hybrid DFT, the ExxSS method with an appropriately chosen auxiliary function achieves substantial improvements not only in ground-state energies but also in other bulk properties. Compared to the previously developed Staggered Mesh Method for Fock exchange, which enhances the asymptotic convergence of the FSE via staggered $\vk$-grids for occupied orbitals \cite{quitonStaggeredMeshMethod2024}, the ExxSS method eliminates the need to compute densities on twice as many $\vk$-points and incurs minimal additional cost relative to standard SCF calculations.

The effectiveness of ExxSS with the newly proposed auxiliary functions arises from the observation that the structure factor can be more accurately approximated by non-Gaussian functions, particularly at low to moderate $|\vq + \vG|$, that decay more slowly and incorporate exponential or even algebraic behavior. Among these, the Quartic Exponential and Gaussian-Lorentzian auxiliary functions demonstrate the best performance in recovering the thermodynamic limit (TDL) from sparse $\vk$-meshes, with their effectiveness persisting across larger basis sets. Although this work focuses on the PBE0 functional, similar improvements are expected for other hybrid functionals with comparable or greater exact exchange fractions. We recommend the Quartic Exponential auxiliary function for general use with ExxSS due to its excellent accuracy, low parameter count, and insensitivity to the cutoff parameter $q_\mathrm{cut}$.

While ExxSS already significantly improves the finite-size error behavior for a broad range of materials, it can also be extended to higher-order correlated methods such as MP$n$ and coupled cluster. In these cases, the structure factor should be replaced by that obtained from higher-level theories~\cite{mihmHowExchangeEnergy2023,schaferSamplingReciprocalCoulomb2024}. We also note that in such methods, the FSE in the structure factor can be comparable in magnitude to the finite-size error in the final correlation energy itself \cite{xingUnifiedAnalysisFinitesize2024,xingInverseVolumeScaling2024}.

\begin{acknowledgement}

This material is based upon work supported by the National Science Foundation Graduate Research Fellowship Program under Grant No. DGE 2146752 (SJQ) and the U.S. Department of Energy, Office of Science, Office of Advanced Scientific Computing Research and Office of Basic Energy Sciences, Scientific Discovery through the Advanced Computing (SciDAC) program under Award Number DE-SC0022364 (LL, MHG). LL is a Simons Investigator in Mathematics. This research also used the Savio computational cluster resource provided by the Berkeley Research Computing program at the University of California, Berkeley (supported by the UC Berkeley Chancellor, Vice Chancellor for Research, and Chief Information Officer). The authors would also like to acknowledge Zhen Huang for helpful discussions.

\end{acknowledgement}

\begin{suppinfo}

Raw data for exchange energies, total energies, and Birch-Murnaghan fit data, for the systems discussed in the work; further results for lonsdaleite; Lorentzian auxiliary function performance; and a comparison of finite-difference and least-squares fitting $\sigma$ values for the Gaussian auxiliary function. This information is available free of charge via the Internet at \url{http://pubs.acs.org}.

\end{suppinfo}

\bibliography{SS1G-0125_abbrev}

\clearpage


\section*{Supplementary material}

# 1 Auxiliary Function Performance

## 1.1 Silicon

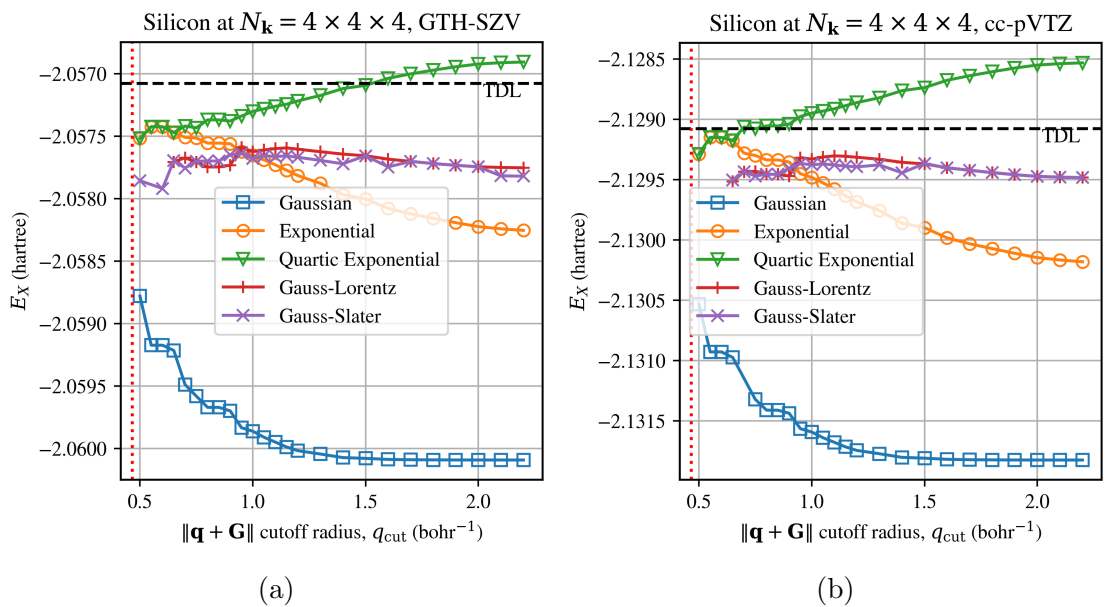

(a)

(b)

Figure S1: Performance of different auxiliary functions with the PBE0 functional at different basis sets for Silicon. The TDL was computed by taking the Ewald/Madelung curve and fitting an $N_{\mathbf{k}}^{-1}$ power law polynomial to the exchange energy values at $N_{\mathbf{k}} = [6^3, 8^3]$ and $[4^3, 6^3]$ for GTH-SZV and cc-pVTZ respectively. The dashed vertical red line indicates the norm of the $\mathbf{q}$-point closest to the origin

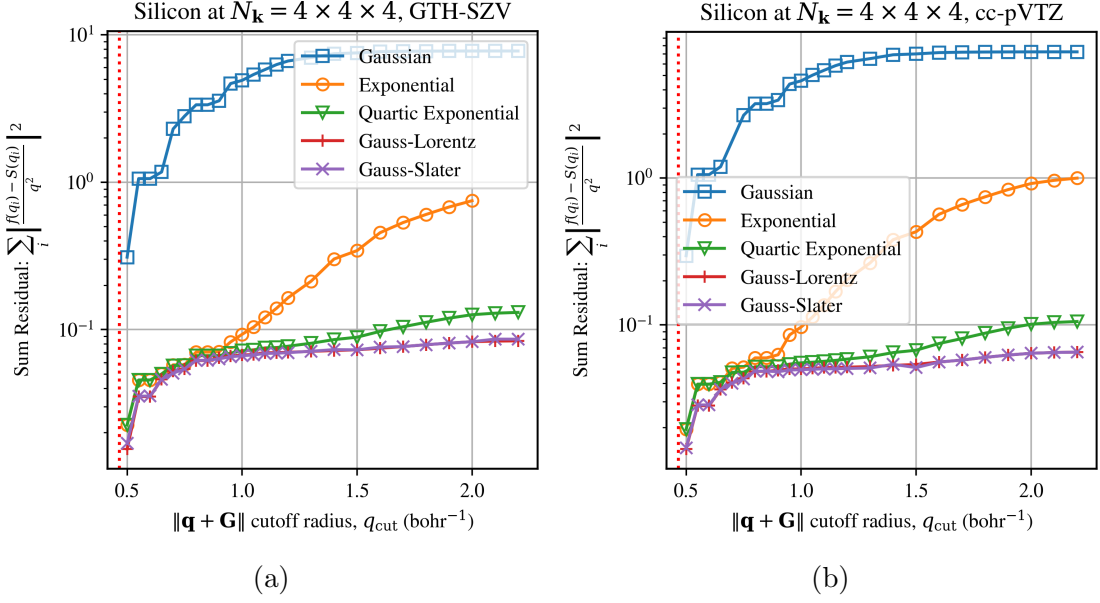

Figure S2: Residual values of each auxiliary function as a function of $q_{cut}$ for Silicon. The dashed vertical red line indicates the norm of the **q**-point closest to the origin

## 1.2 Lorentzian Performance

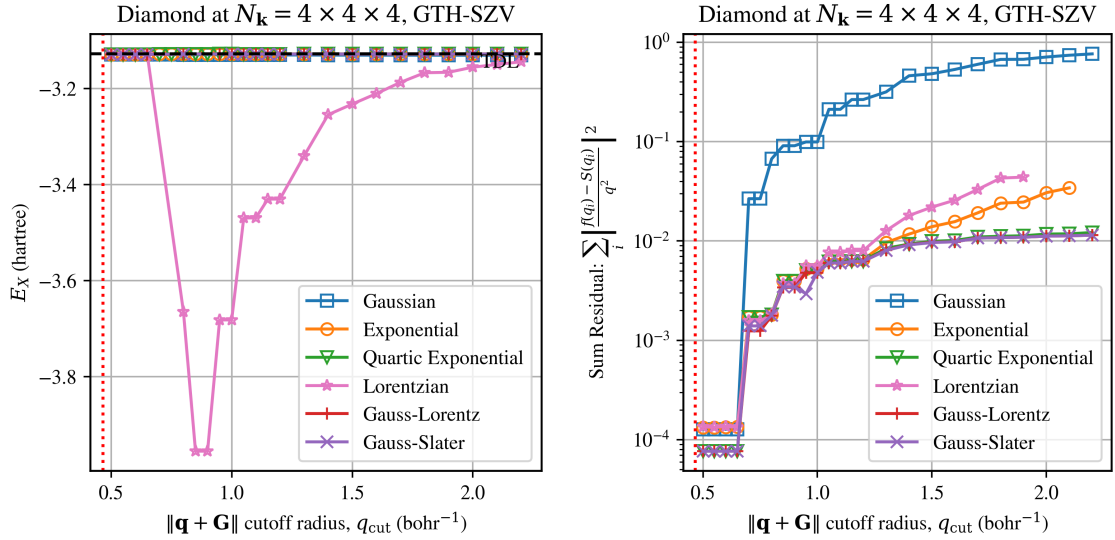

Figure S3: Performance of the Lorentzian auxiliary function for Diamond at $N_{\mathbf{k}} = 4 \times 4 \times 4$ at PBE0/GTH-SZV.

## 2 Lonsdaleite with an unshifted k-mesh

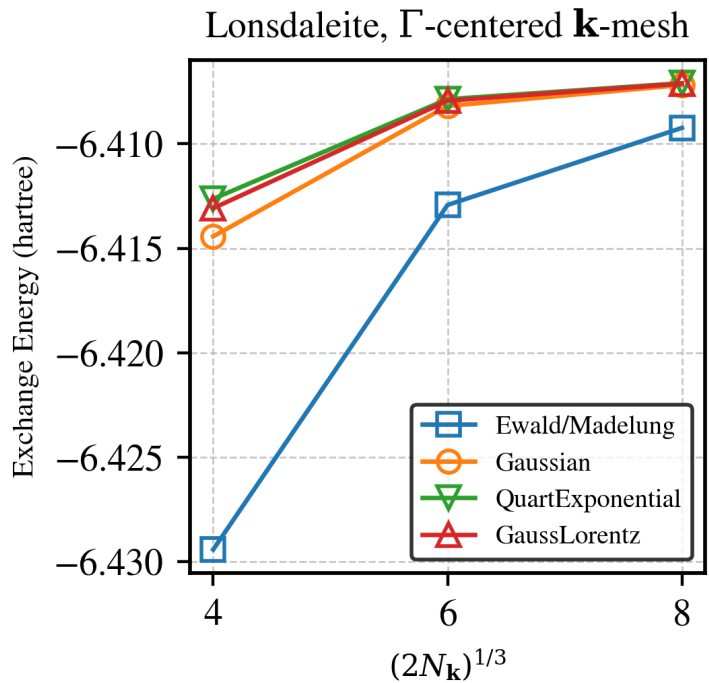

Figure S4: Exchange energy FSE curves for lonsdaleite at the PBE0/GTH-DZVP level of theory using a $\Gamma$-centered **k**-point mesh

## 3 Comparing $\sigma$ values obtained from least-squares fitting and finite differences

In addition to fitting a Gaussian to the structure factor using a low $q_{\text{cut}}$, one could also directly compute the curvature of the structure factor at the $\Gamma$-point through finite differences. This is very similar to the approach that Duchemin and Gygi have taken, only that they apply a correction to the structure factor at each **k** and $i$ by taking the analytical integral of a paraboloid with the computed finite-difference curvature.[S1] Here, we only need to compute

the curvature for $S(\mathbf{q})$ at the $\Gamma$-point.

$$S_i^{(2)}(\mathbf{0}) \approx \frac{S(\Delta\mathbf{q}_i) - 2S(\mathbf{0}) + S(-\Delta\mathbf{q}_i)}{2|\Delta\mathbf{q}_i|^2}, \tag{1}$$

where $\Delta\mathbf{q}_i = \mathbf{b}_i/N_\mathbf{k}$, and $\{\mathbf{b}_i\}$ are the reciprocal lattice vectors. Under this approximation, we expect that the error of the curvature would be $\mathcal{O}((\Delta\mathbf{q})^2)$. We take the overall curvature $S^{(2)}(\mathbf{0})$ to be the average of the curvatures in each direction (though we note that the curvatures for diamond and silicon are isotropic), and then set the Gaussian width $\sigma = \sqrt{N_{\mathrm{occ}}/2S^{(2)}(\mathbf{0})}$. In Figure S5, we compare the $\sigma$ values obtained from finite differences to those from fitting a Gaussian to the points closest to the origin and find that the latter consistently provides lower $\sigma$ values. Then, in Figure S6, the $\sigma$ values from each method

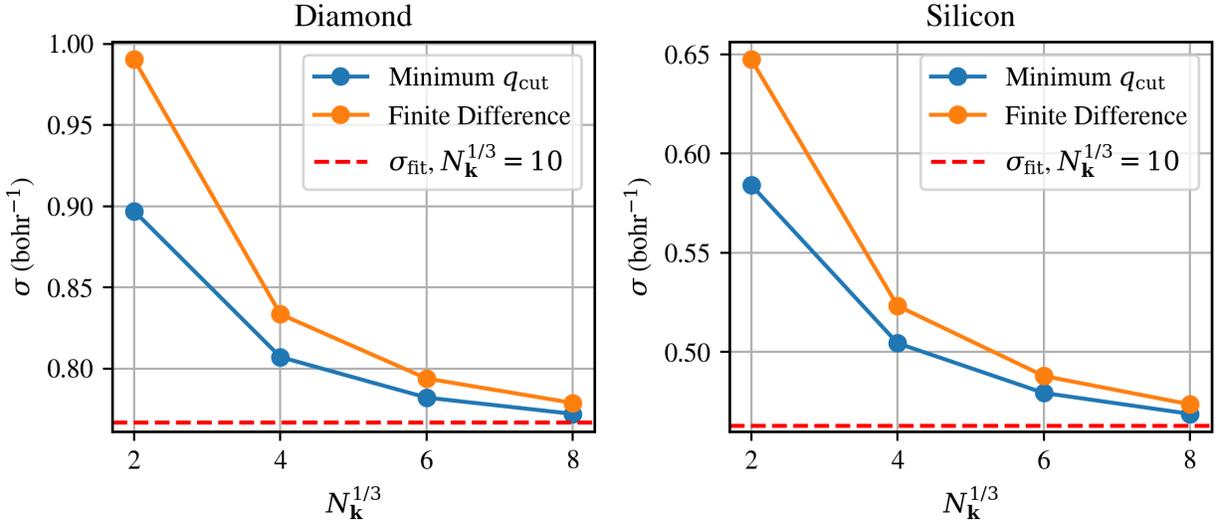

Figure S5: Comparison of the $\sigma$ values obtained from finite difference versus least squares fitting (using the minimum $q_{\mathrm{cut}}$) procedures. We use the least-squares $\sigma_{\mathrm{fit}}$ from a $10 \times 10 \times 10$ calculation to estimate $\sigma$ value that recovers the curvature of the structure factor, shown with the red dashed line. Calculations were at the PBE0/SZV-GTH level of theory.

is used to compute the corrected exchange energy, showing that the two methods provide similar values but the Gaussian-fitting provides slightly more accurate values, particularly at smaller k-mesh sizes. Therefore, all results in this work using the Gaussian auxiliary function will use the least-squares fitting procedure to determine $\sigma$.

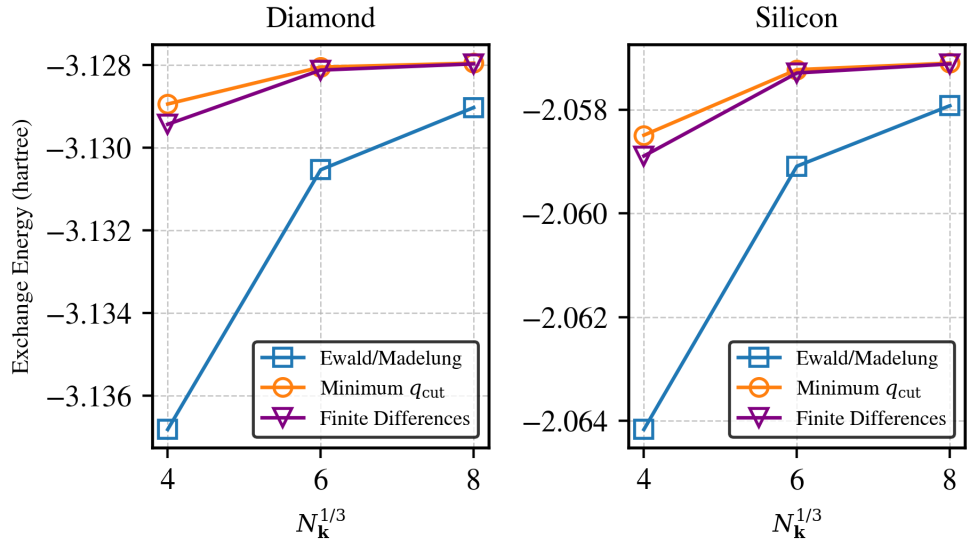

Figure S6: Finite-size error curves resulting from the fitting and finite difference approaches.

**TOC Graphic**

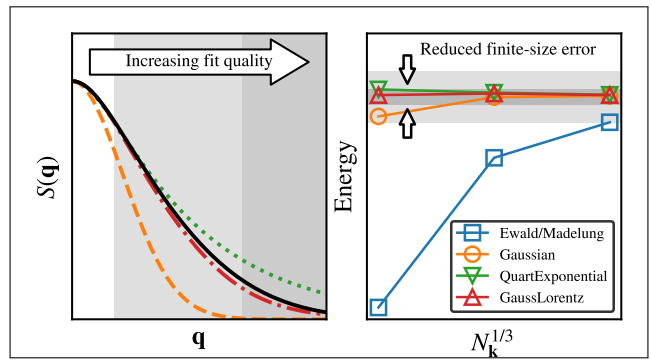



\end{document}